\documentclass[reprint,aps,prl,floatfix,
tightenlines]{revtex4-2}
\usepackage{chemformula}
\usepackage[T1]{fontenc}
\usepackage{booktabs}
\usepackage{upgreek}
\usepackage{silence}
\usepackage{lineno}
\usepackage[colorlinks=true, citecolor=magenta, linkcolor=blue, filecolor=magenta, urlcolor=magenta]{hyperref}

\begin{document}

\title
  {Buried unstrained germanium channels: a lattice-matched platform for quantum technology}
\author{Davide Costa}
\affiliation{QuTech and Kavli Institute of Nanoscience, Delft University of Technology, Lorentzweg 1, 2628 CJ Delft, Netherlands}
\author{Patrick Del Vecchio}
\affiliation{QuTech and Kavli Institute of Nanoscience, Delft University of Technology, Lorentzweg 1, 2628 CJ Delft, Netherlands}
\author{Karina Hudson}
\affiliation{QuTech and Kavli Institute of Nanoscience, Delft University of Technology, Lorentzweg 1, 2628 CJ Delft, Netherlands}
\author{Lucas E. A. Stehouwer}
\affiliation{QuTech and Kavli Institute of Nanoscience, Delft University of Technology, Lorentzweg 1, 2628 CJ Delft, Netherlands}
\author{Alberto Tosato}
\affiliation{QuTech and Kavli Institute of Nanoscience, Delft University of Technology, Lorentzweg 1, 2628 CJ Delft, Netherlands}
\author{Davide Degli Esposti}
\affiliation{QuTech and Kavli Institute of Nanoscience, Delft University of Technology, Lorentzweg 1, 2628 CJ Delft, Netherlands}
\author{Vladimir Calvi}
\affiliation{QuTech and Kavli Institute of Nanoscience, Delft University of Technology, Lorentzweg 1, 2628 CJ Delft, Netherlands}
\author{Luca Moreschini}
\affiliation{QuTech and Kavli Institute of Nanoscience, Delft University of Technology, Lorentzweg 1, 2628 CJ Delft, Netherlands}
\author{Mario Lodari}
\affiliation{QuTech and Kavli Institute of Nanoscience, Delft University of Technology, Lorentzweg 1, 2628 CJ Delft, Netherlands}
\author{Stefano Bosco}
\affiliation{QuTech and Kavli Institute of Nanoscience, Delft University of Technology, Lorentzweg 1, 2628 CJ Delft, Netherlands}
\author{Giordano Scappucci}
\email{g.scappucci@tudelft.nl}
\affiliation{QuTech and Kavli Institute of Nanoscience, Delft University of Technology, Lorentzweg 1, 2628 CJ Delft, Netherlands}

\date{\today}

\begin{abstract}

Strained germanium ($\varepsilon$-Ge) and strained silicon ($\varepsilon$-Si) buried quantum wells have enabled advanced spin-qubit quantum processors. However, in the absence of suitable lattice-matched substrates, $\varepsilon$-Ge and $\varepsilon$-Si are deposited on defective, metamorphic SiGe buffers, which may impact device performance and scaling. Here an alternative platform is introduced based on the heterojunction between bulk unstrained Ge and a lattice-matched strained silicon-germanium ($\varepsilon$-SiGe) barrier, eliminating the need for metamorphic buffers altogether. In a structure with a 52-nm-thick $\varepsilon$-SiGe barrier, a low-disorder two-dimensional hole gas is demonstrated with a high-mobility of $1.33 \times 10^{5} ~\mathrm{cm^2/Vs}$ and a low percolation density of $1.4(1)\times 10^{10} ~\mathrm{cm^{-2}}$. Quantum transport shows that holes confined in the buried unstrained Ge channel have a strong density-dependent in-plane effective mass and out-of-plane $g$-factor, pointing to a significant heavy-hole--light-hole mixing in agreement with theory. Measurements of Zeeman-split levels in quantum point contacts further highlight this character, showing a two-fold larger in-plane $g$-factor in Ge than in $\varepsilon$-Ge. The prospects of strong spin–orbit interaction, isotopic purification, and of hosting superconducting pairing correlations make this platform appealing for fast quantum hardware and hybrid quantum systems.
\end{abstract}

\maketitle

\section{Introduction}

Continuous advances in materials underpin the development of semiconductor quantum technology \cite{de_leon_materials_2021} based on spin qubits in quantum dots \cite{burkard2023semiconductor} and superconductor–semiconductor hybrid devices \cite{prada_andreev_2020}.
Spin qubits were first realized in GaAs-based heterostructures \cite{petta_coherent_2005, koppens_driven_2006}, where lattice-matched GaAs/AlGaAs epitaxy produced buried, high-mobility electron gases and electrostatically defined quantum dots largely free of disorder \cite{hanson_spins_2007}. However, the hyperfine interaction with the abundant nuclear spins in III–V materials strongly limited spin coherence \cite{cywinski_electron_2009}, motivating a shift toward group-IV semiconductors Si and Ge, which have a low natural abundance of nuclear spins and can be further isotopically purified \cite{itoh_high_1993,saraiva_materials_2022, moutanabbir_nuclear_2024}. In Si metal-oxide-semiconductor devices (Si-MOS), isotopically purified Si epilayers are lattice matched to pristine, Si substrates \cite{fukatsu_effect_2003,sabbagh_quantum_2019} and long spin coherence times have been demonstrated \cite{veldhorst2014}, while maintaining compatibility with advanced semiconductor manufacturing \cite{zwerver2022qubits, steinacker_industry-compatible_2025}. Yet, qubits in Si-MOS are defined at the semiconductor-oxide interface, introducing electrostatic disorder and charge noise and posing a challenge for scaling \cite{cifuentes_bounds_2024}.

Alternatively, spin-qubits in strained Ge ($\varepsilon$-Ge)  \cite{hendrickx_four-qubit_2021,hendrickx_sweet-spot_2024,stehouwer_exploiting_2025} and strained Si ($\varepsilon$-Si) \cite{yoneda_quantum-dot_2018,xue_quantum_2022,noiri_fast_2022,neyens_probing_2024} buried quantum wells \cite{scappucci_germanium_2021,scappucci_crystalline_2021} may experience a quiet electrical environment because the noisy semiconductor-oxide interface is separated by an epitaxial SiGe barrier \cite{paquelet_wuetz_reducing_2023}. 
In the absence of high-quality SiGe wafers for epitaxy, $\varepsilon$-Ge and $\varepsilon$-Si quantum wells are grown on strain-relaxed SiGe buffers, which act as metamorphic substrates \cite{deelman_metamorphic_2016} bridging the lattice mismatch with the underlying Ge or Si wafers. 
However, these SiGe metamorphic substrates rely on networks of dislocations for strain-release and are inherently defective, introducing topographic, strain, chemical, and band offset fluctuations in the strained quantum wells \cite{evans_nanoscale_2012,corley-wiciak_nanoscale_2023,corley-wiciak_lattice_2023}, thereby challenging the performance and cross-wafer uniformity of quantum devices.

Here, we develop a group IV semiconductor heterostructure that has the potential to unite in a single material stack three key merits sought for spin qubits materials---buried channels for low electrostatic disorder, lattice matching to the substrate for a defect-free crystal, and possibility of isotopic purification for long spin coherence—whereas preceding architectures offered only subsets of these advantages.
The heterostructure is based on the heterojunction between unstrained Ge and a strained SiGe ($\varepsilon$-SiGe) barrier that is lattice-matched to a pristine Ge substrate, eliminating the need for metamorphic substrates.
Building on the recent use of Ge wafers for SiGe heterostructures epitaxy \cite{stehouwer2023germanium,stehouwer_exploiting_2025}, this approach realizes a seminal but long-overlooked design principle \cite{people_indirect_1986}: that two-dimensional systems can be formed in elemental Ge by exploiting the band alignment of coherently strained SiGe barriers on Ge substrates.

These early oversimplified calculations \cite{people_indirect_1986} neglected the significant energy splitting between heavy-holes (HH) and light holes (LH) due to quantum confinement \cite{Winkler1996,Winkler2003,scappucci_germanium_2021} in the Ge channel at the heterojunction, leading to the challenging proposal of depositing highly strained Si$_{0.5}$Ge$_{0.5}$ barriers to achieve sufficient band offset for confining a two-dimensional hole gas (2DHG). This approach proved impractical in early experiments \cite{wagner_observation_1989} and was soon abandoned in favour of $\varepsilon$-Ge quantum wells on strain-relaxed SiGe buffers \cite{murakami_high_1990}.
Instead, from measurements on undoped insulated-gate field-effect transistors, supported by comprehensive self-consistent Poisson--Schr\"odinger simulations, we demonstrate that even a moderately tensile-strained Si$_{0.2}$Ge$_{0.8}$ barrier provides sufficient band offset for robust confinement of a 2DHG at the buried heterojunction in Ge. The 2DHG has high mobility, low percolation density, and shows fractional quantum Hall states at low density.
Quantum transport, supported by theoretical calculations, reveals electrically-tunable in-plane effective mass ($m^*$) and out-of-plane $g$-factor ($g^*_{\perp}$), highlighting confinement dominated moderate HH--LH energy splitting leading to significant HH--LH mixing and enhanced spin--orbit interaction in unstrained Ge that marks a clear distinction from strain-dominated HH-LH large splittings in $\varepsilon-$Ge quantum wells. This distinction is reinforced by further confinement into quantum point contacts, where the characterization of Zeeman-split one dimensional subbands reveals a much larger in-plane $g$-factor ($g^*_{\parallel}$) in Ge compared to $\varepsilon$-Ge.

\begin{figure*}[t]
  \centering
	\includegraphics[width=170mm]{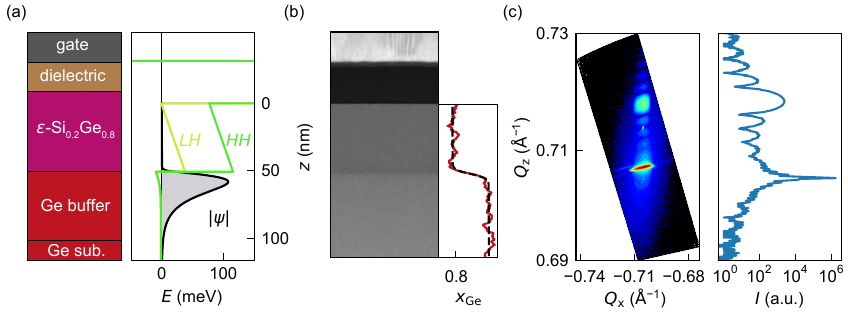}
	\caption{(a) Layer schematic of the semiconductor heterostructure and gate stack featuring an heterojunction between an unstrained Ge buffer and a strained SiGe ($\varepsilon$-SiGe) barrier (left) and simulated band-edges (right) with heavy holes (HH, green line), light holes (LH, light green line). The ground state heavy hole wavefunction $|\psi|$ (black line) resides primarily in the Ge buffer, lattice-matched to the Ge substrate. The Fermi energy is set as the reference energy at $0 ~\mathrm{eV}$. (b) HAADF-STEM image of the active layers of the Ge/$\varepsilon$-SiGe heterostructure (left) with EDX profile (red line, right) showing the Ge alloy concentration ($x_{\mathrm{Ge}}$) as a red curve and fit to a sigmoid function (dotted black line). The vertical $z$-axis scale is as in (a). (c) X-ray diffraction reciprocal space map of the (-404) planes (left) as a function of the in-plane ($Q_{\mathrm{x}}$) and out-of-plane ($Q_{\mathrm{z}}$) inverse of lattice spacing, with a high-resolution $\omega$/$2\theta$ scan around the Ge (004) peak (right). $I$ is the signal intensity in arbitrary units.}
\label{fig:one}
\end{figure*}

\section{Results and discussion}
The lattice-matched Ge/$\varepsilon$-SiGe heterostructure is grown by reduced-pressure chemical vapour deposition on a $100~\mathrm{mm}$ Ge(001)~wafer. 
As shown in Fig.~\ref{fig:one}(a) (left panel), the semiconductor stack design comprises an unstrained $250~\mathrm{nm}$ epitaxial Ge buffer layer, a tensile-strained $52 ~\mathrm{nm}$ Si$_{0.2}$Ge$_{0.8}$ barrier, and a final sacrificial Si cap.
Details of the epitaxy conditions for Ge and SiGe layers on Ge wafers are reported in \cite{stehouwer2023germanium}.
Hall-bar shaped heterostructure field effect transistors (H-FETs) are fabricated with a low-thermal budget process featuring platinum-germanosilicide ohmic contacts and an Al$_{2}$O$_{3}$/Ti/Pd gate stack as described in \cite{sammak_shallow_2019,lodari_low_2021}. 
Unlike defective metamorphic substrates, where strain relaxation is promoted by pre-existing dislocations\cite{matthews_defects_1974}, growth on a pristine substrate allows for a sufficiently thick strained barrier to separate the heterojunction from the disordered dielectric, while still remaining below the theoretical critical thickness for strain relaxation \cite{people_calculation_1985,bean_strained-layer_1985,alam_critical_2019}.

One-dimensional Schr\"{o}dinger–Poisson simulations of the heavy-hole (HH) and light-hole (LH) band edges along the growth direction $z$ are shown in the right panel of Fig.~\ref{fig:one}(a).
The electric field from the insulated top-gate induces a triangular quantum well at the buried Ge/$\varepsilon$-Si$_{0.2}$Ge$_{0.8}$ heterojunction for accumulation of a 2DHG \cite{Scappucci_patent}, advancing the theoretical understanding of these heterojunction presented in earlier work \cite{people_indirect_1986}.
The HH wavefunction ($|\psi|$) resides predominantly in the unstrained Ge layer, where charge carrier confinement is promoted by a band-offset of about $125 ~\mathrm{meV}$ at the heterojunction, arising from the strain-induced splitting of the HH and LH bands in the $\varepsilon$-Si$_{0.2}$Ge$_{0.8}$ layer and from quantum confinement of gate-induced charge within the Ge layer.
While the band offset is comparable to that in $\varepsilon$-Ge quantum wells ($\sim130 ~\mathrm{meV}$) \cite{sammak_shallow_2019}, the HH--LH energy splitting is quite different.
In this case, quantum confinement in the unstrained Ge layer yields a HH--LH splitting of about $3 ~\mathrm{meV}$ --- much smaller than the $70 ~\mathrm{meV}$ typically observed in $\varepsilon$-Ge.
Nevertheless, this splitting remains sufficiently large to avoid the valley splitting challenge present for electrons in strained Si quantum wells \cite{friesen_magnetic_2006,degliesposti2024low}.

Figure~\ref{fig:one}b shows a high angle annular dark field (HAADF) scanning transmission electron microscopy (STEM) image of the active layers of the heterostructure, along with the energy dispersive X-ray (EDX) profile of the Ge concentration $x_{\mathrm{Ge}}$.
The image confirms the high-quality epitaxial deposition of a $52(1) ~\mathrm{nm}$ thick Si$_{0.2}$Ge$_{0.8}$ barrier with no visible defects crossing the buried heterojunction.
We estimate an upper bound for the characteristic length-scale $4\tau$ of the heterojunction interface of $3.8(3) ~\mathrm{nm}$ by fitting the Ge content profile to a sigmoid model (see the Supporting Information).

As shown in the Supporting Information, characterisation of the as-grown heterostructure by atomic force microscopy and scanning Raman spectroscopy indicates that the Si$_{0.2}$Ge$_{0.8}$ barrier is flat (root mean square roughness $\sim 0.4 ~\mathrm{nm}$), tensile-strained (average in-plane strain $\overline{\varepsilon_{\parallel}} = 1.0(4) \times 10^{-2}$), and exhibits no signs of a cross-hatch pattern \cite{zoellner_imaging_2015}.
This marks a major difference compared to $\varepsilon$-Ge (or $\varepsilon$-Si) quantum wells, where the strain field associated with the underlying network of misfit dislocations in the strain-relaxed buffer induces a prominent cross-hatch pattern \cite{stehouwer2023germanium,corley-wiciak_nanoscale_2023,degliesposti2024low}.

\begin{figure*}[t]
  \centering
	\includegraphics[width=170mm]{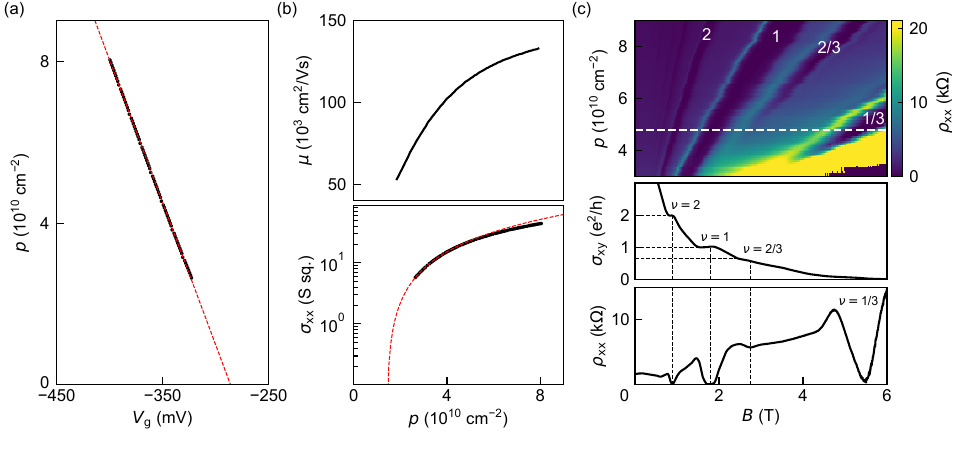}
	\caption{(a) Hall density ($p$) as a function of gate voltage ($V_{\mathrm{g}}$) for a Ge/$\varepsilon$-SiGe heterostructure field effect transistor (black curve) and corresponding linear fit (dashed red line). (b) Hole mobility ($\mu$) in the top panel and longitudinal conductivity ($\sigma_{\mathrm{xx}}$) as black curve in the bottom panel as a function of $p$. The red dashed line is a fit to percolation theory in two-dimensions. (c) Top panel: Landau fan diagram with longitudinal resistivity ($\rho_{\mathrm{xx}}$) as a function of perpendicular magnetic field $B$ and $p$, obtained by stepping $B$ and sweeping $V_{\mathrm{g}}$. The dashed white line marks the density $p = 4.8 \times 10^{10} ~\mathrm{cm^{-2}}$ for detailed measurements of transversal conductivity $\sigma_{\mathrm{xy}}$ (central panel) and of $\rho_{\mathrm{xx}}$ (bottom panel). Quantum Hall plateaus and related Shubnikov--de Haas oscillation minima for integer and fractional states are highlighted with black dashed lines. All measurements are performed at a temperature of $60 ~\mathrm{mK}$ measured at the mixing chamber of the dilution refrigerator}
\label{fig:two}
\end{figure*}

In Fig.~\ref{fig:one}(c) (left panel), high resolution X-ray diffraction reciprocal space mapping using the (-404) reflection shows that the $\varepsilon$-Si$_{0.2}$Ge$_{0.8}$ and Ge peaks lie on the same vertical line.
The position $Q_{\mathrm{x}}$ of their lattice spacing in reciprocal space differs by only 0.07\%, highlighting the similar in-plane lattice constant and confirming the heterostructure is lattice-matched.
In the $\omega$-$2\theta$ scan around the Ge (004) peak (right panel), pronounced Pendell\"{o}sung fringes indicate high crystalline quality with flat, parallel interfaces \cite{bowen1998high}.
Analysis of their separation yields a $263.1(1)~\mathrm{nm}$ epitaxial Ge layer with a $52.7(5) ~\mathrm{nm}$ $\varepsilon$-Si$_{0.2}$Ge$_{0.8}$ barrier on top, in agreement with the intended design and HAADF-STEM characterisation.

The electrical properties of the buried Ge/$\varepsilon$-SiGe heterojunction are characterized by magnetotransport measurements of the H-FET at a temperature of $60~\mathrm{mK}$, using four-terminal low-frequency lock-in techniques. 
Applying a negative gate voltage $V_{\mathrm{g}}$ forms a 2DHG in accumulation mode with a tunable carrier density $p$.
In the Supporting Information we show the two-terminal turn-on curve of the H-FET, measuring the source-drain current as a function of $V_{\mathrm{g}}$.
The observed linear $p$-$V_{\mathrm{g}}$ relationship in Fig.~\ref{fig:two}(a) (black curve) confirms a capacitively induced channel and excludes charge tunnelling into the SiGe LH states or towards the surface \cite{su_effects_2017}.
However, applying increasingly negative gate voltages above a density of $8.0 \times 10^{10} ~\mathrm{cm^{-2}}$ causes a shift in the device characteristics due to charge trapping within the dielectric or at the semiconductor-dielectric interface \cite{lodari_lightly_2022, massai_impact_2024}, screening the further charge accumulation at the buried interface.
From the fit (dashed red line) we estimate a capacitance per unit area $C$ of $112.87(1)~\mathrm{nF/cm^2}$, in agreement with $\varepsilon$-Ge quantum wells with similar barrier and dielectric thicknesses \cite{lodari_low_2021}, indicating the 2DHG is formed at the buried heterojunction. 
Furthermore, we measure a minimum Hall density of $2.6 \times 10^{10} ~\mathrm{cm^{-2}}$, on par with $\varepsilon$-Ge quantum wells used for large spin qubit arrays \cite{lodari_low_2021,stehouwer2023germanium}, hinting at a very low disorder channel.

The top and bottom panels in Fig.~\ref{fig:two}(b) show the density-dependent hole mobility $\mu(p)$ and longitudinal conductivity $\sigma_{\mathrm{xx}}(p)$, respectively.
We measure a maximum mobility $\mu_{\mathrm{max}}$ of $1.33 \times 10^{5} ~\mathrm{cm^2/Vs}$ at a saturation density $p_{\mathrm{sat}}$ of $8.0 \times 10^{10} ~\mathrm{cm^{-2}}$.
Fitting the density-dependent conductivity to 2D percolation theory, $\sigma_{\mathrm{xx}} \propto (p-p_{\mathrm{p}})^{1.31}$ \cite{tracy_observation_2009,fogelholm1980conductivity}, we estimate a percolation-induced critical density $p_{\mathrm{p}}$ of $1.4(1) \times 10^{10} ~\mathrm{cm^{-2}}$, approaching the value of $1.22(3) \times 10^{10} ~\mathrm{cm^{-2}}$ achieved in 
$\varepsilon$-Ge/SiGe quantum wells grown on Ge wafers with a similarly thick SiGe barrier \cite{stehouwer2023germanium}. This comparison suggests a similarly low-disorder potential landscape at low densities, implying that quantum dots of about $1/\sqrt{p_\mathrm{p}} \sim 80~\mathrm{nm}$ in size, informative about the average distance between charge traps, are essentially disorder-free~\cite{scappucci_crystalline_2021}.
However, the maximum mobility in Ge/$\varepsilon$-SiGe is more than an order of magnitude lower than in $\varepsilon$-Ge/SiGe. We speculate that the discrepancy in mobility at high density arises from impurity scattering from unwanted oxygen accumulation at the Ge/$\varepsilon$-SiGe interface \cite{mi_magnetotransport_2015,lu2026impact}, as shown by the secondary ion mass spectrometry in the Supporting Information, and from interface roughness scattering \cite{costa_reducing_2024} associated with the rather diffused Ge/$\varepsilon$-SiGe interface. 
Starting from this proof-of-principle heterostructure, we expect to reduce oxygen incorporation in the Ge and SiGe films by installing chemical filters in the gas precursor lines, leading to a potential mobility improvement up to $4\times$ \cite{lu2026impact}, or by refining the growth temperature profile during epitaxy \cite{bedell_invited_2020,myronov_holes_2023} .
Furthermore, as discussed below, the heavier mass associated with HH--LH mixing at the higher end of the investigated density range contributes significantly to the observed mobility difference with $\varepsilon$-Ge/SiGe quantum wells.
A supplementary comparison of mobility, percolation density, and transport scattering time across group-IV platforms for spin qubits is provided in the Supporting Information. This comparison shows that Ge/$\varepsilon$-SiGe already significantly outperforms $\varepsilon$-Si/SiGe and Si-MOS when benchmarked in the low carrier density regime ($<1 \times 10^{11} ~\mathrm{cm^{-2}}$) relevant for quantum dot qubit operation.

\begin{figure*}[t]
  \centering
	\includegraphics[width=170mm]{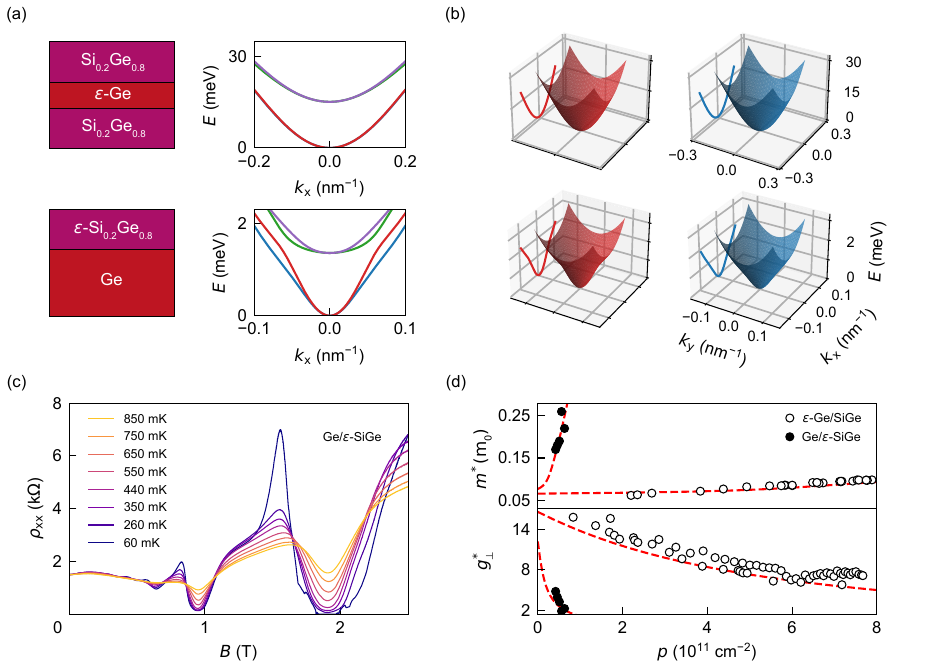}
	\caption{(a) Simulation of the first four energy levels at zero magnetic field and $k_{\mathrm{y}}=0$ for the 2DHG in the strained Ge ($\varepsilon$-Ge) quantum well (top, four HH levels) and in the unstrained Ge at the Ge/$\varepsilon$-SiGe strained barrier heterojunction (bottom, two HH levels, blue and red, and two LH levels, green and purple). (b) Corresponding simulated dispersion relation of the spin up (red) and spin down (blue) ground state of the $\varepsilon$-Ge quantum well (top) and of the unstrained Ge channel in the Ge/$\varepsilon$-SiGe heterojunction (bottom). (c) Longitudinal resistivity ($\rho_{\mathrm{xx}}$) as a function of perpendicular magnetic field $B$ at Hall density $p = 4.55 \times 10^{10} ~\mathrm{cm^{-2}}$, measured at different temperatures, ranging from from $60 ~\mathrm{mK}$ (blue) to $850 ~\mathrm{mK}$ (yellow) and measured at the mixing chamber of the dilution refrigerator. (d) Extracted in-plane effective mass ($m^*$) and effective out-of-plane $g$-factor ($g^*_{\perp}$) values for the 2DHG in Ge/$\varepsilon$-SiGe (filled circles) and in $\varepsilon$-Ge/SiGe (open circles) with theoretical simulation of these parameters (dashed red lines).}
\label{fig:three}
\end{figure*}

We further highlight the low-disorder properties of the 2DHG by performing quantum transport measurements at higher perpendicular magnetic fields. 
The Landau level fan diagram in the top panel of Fig.~\ref{fig:two}(c) shows $\rho_{\mathrm{xx}}$ as a function of perpendicular $B$ and $p$. 
This has been calculated from the measurement of $\rho_{\mathrm{xx}}$ of a function of sweeping $V_{\mathrm{g}}$ and stepping perpendicular $B$ as shown in the Supporting Information.
Dark blue regions correspond to dips in $\rho_{\mathrm{xx}}$ and highlight the density-dependent evolution of integer and fractional filling factors $\nu = 1/3, 2/3, 1, 2$, which fan out toward higher magnetic field and density.
The dashed white line in the fan diagram indicates the magnetic field range selected for higher resolution measurements of $\rho_{\mathrm{xx}}$ and the transversal conductivity $\sigma_{\mathrm{xy}}$ at a fixed density of $p = 4.8 \times 10^{10} ~\mathrm{cm^{-2}}$, as shown in bottom and central panels of Fig.~\ref{fig:two}(c), respectively.
A highlight of these measurements is the clear dip in $\rho_{\mathrm{xx}}$ corresponding to $\nu = 1/3$, a fractional quantum Hall state previously observed in lightly-strained Ge quantum wells with hole mobility exceeding one million cm$^2$/Vs \cite{lodari_lightly_2022} and relevant to the direct observation of anionic braiding statistics in GaAs \cite{nakamura_direct_2020}.

We simulate the band structure of the Ge/$\varepsilon$-SiGe strained-barrier heterojunction and, as a reference, of the $\varepsilon$-Ge quantum well including electric and magnetic fields (see Supporting Information) to evaluate and benchmark $m^*$ and $g^*_{\perp}$. 
These band structure parameters exhibit substantial variations between the two systems because of the large difference in HH--LH splitting. 
The simulated spin-dependent energy dispersions at zero magnetic field are shown in Fig.~\ref{fig:three}(a)-(b).

As a reference, in $\varepsilon$-Ge quantum wells, the HH--LH splitting is largely dominated by the compressive strain in Ge, which shifts the lowest LH level roughly $70 ~\mathrm{meV}$ above the HH ground state. 
This large separation leads to an HH energy dispersion that at low densities is mostly parabolic, spin-independent, and with a small in-plane effective mass \cite{lodari_light_2019,terrazos_theory_2021}. 
In contrast, in the unstrained Ge channel at the Ge/$\varepsilon$-SiGe heterojunction, the HH-LH energy splitting is $\sim 3 ~\mathrm{meV}$ and is caused by the electric field-induced quantum confinement, which differs for HHs and LHs because of their different out-of-plane mass. 
In this case, the HH ground state dispersion shows a strong non-parabolicity and spin-dependence at densities comparable to the one measured in our H-FETs ($k_{\mathrm{x}}=0.1 ~\mathrm{nm^{-1}}$ corresponds to $p \sim 10^{11} ~\mathrm{cm^{-2}}$), as seen in Fig.~\ref{fig:three}(a)-(b). 
The large and tunable HH--LH mixing in the ground state of the heterojunction leads to an increase of $m^*$ and a decrease of $g^*_{\perp}$ compared to the strained quantum well, in agreement with the measurements in our devices.

We estimate in-plane $m^*$ and $g^*_{\perp}$ from the temperature-dependent decay of the Shubnikov--de Haas oscillation resistivity $\rho_{\mathrm{xx}}$ minima for different integer filling factors $\nu = p\mathrm{h}/\mathrm{e}B_{\upnu}$, where $B_{\upnu}$ is the magnetic field at integer $\nu$. 
Figure~\ref{fig:three}(c) shows, for the H-FET discussed in Fig.~\ref{fig:two}, an exemplary dataset comprising magnetoresistivity $\rho_{\mathrm{xx}}(B)$ curves measured at a fixed density ($p = 4.55 \times 10^{10} ~\mathrm{cm^{-2}}$) for different temperature $T$  in the $60$ to $850 ~\mathrm{mK}$ range. 
Thermally activated Shubnikov--de Haas oscillations minima are visible at filling factors $\nu = 1,2,3$ from which we extract $m^*$ and $g^*_{\perp}$ according to the procedure in Ref.~\cite{lodari_lightly_2022} and discussed in the Supporting Information. 
We repeat these measurements for five different densities from $4.2$ to $6.3 \times 10^{10} ~\mathrm{cm^{-2}}$ and plot the obtained density dependent $m^*$ and $g^*_{\perp}$ in Fig.~\ref{fig:three}(d) (filled circles). 
At the lowest measured density ($4.2 \times 10^{10} ~\mathrm{cm^{-2}}$) we obtain an effective mass of $0.17\mathrm{m_{\mathrm{0}}}$ and a $g^*_{\perp}$ of $4.85$. 
We also report, as a comparison, previous data from $\varepsilon$-Ge quantum wells \cite{lodari_light_2019, costa_data_2025} (open circles).

\begin{figure*}[t]
  \centering
	\includegraphics[width=170mm]{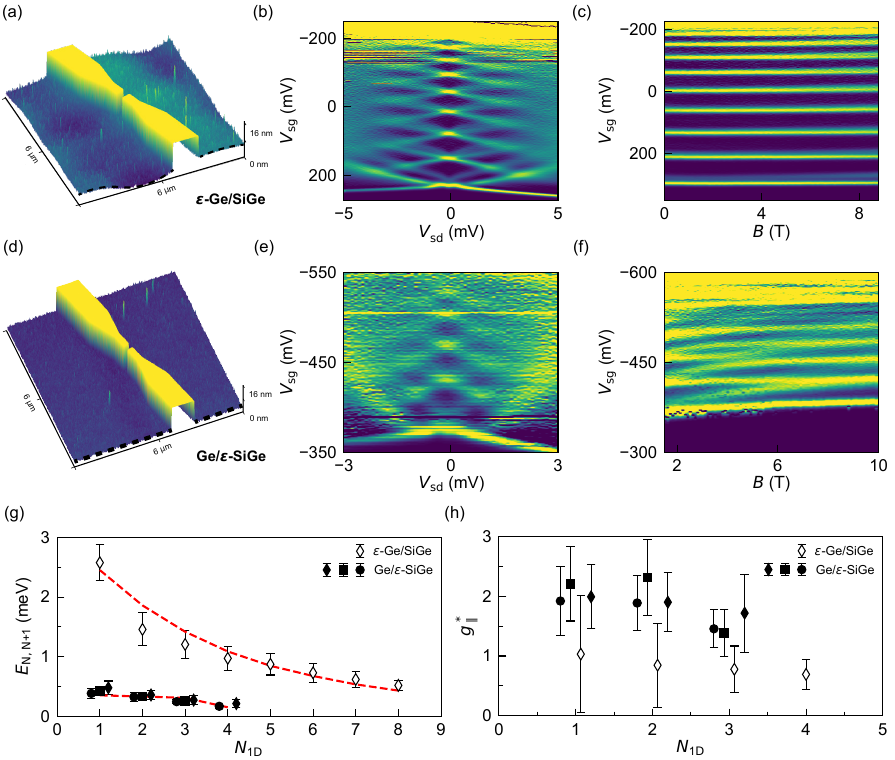}
	\caption{(a),(d) Atomic force microscopy (AFM) images of quantum point contact (QPC) devices fabricated on $\varepsilon$-Ge/SiGe and on Ge/$\varepsilon$-SiGe heterostructures, showing the device constriction side gates and the absence of cross-hatch pattern in the lattice-matched platform. (b),(e) Source–drain bias spectroscopy of the differential transconductance ($\partial G_{\mathrm{xx}}/\partial V_{\mathrm{sg}}$) as a function of side-gate voltage ($V_{\mathrm{sg}}$) and source–drain bias ($V_{\mathrm{sd}}$) at a density of $8\times10^{10}~\mathrm{cm^{-2}}$, revealing clear 1D subband quantization. The asymmetry with respect to zero bias
is due to a DC voltage offset of $\sim 0.1$ mV in our measurement electronics. (c),(f) Zeeman spectroscopy of the QPCs at a density of $8\times10^{10}~\mathrm{cm^{-2}}$, showing the evolution of spin-resolved 1D subbands with in-plane magnetic field $B$. (g) Extracted 1D subband spacings ($E_{N,N+1}$) from (b),(e) as a function of subband index ($N$) for the $\varepsilon$-Ge quantum well (white points, measured at Hall density of $8 \times 10^{10}~\mathrm{cm^{-2}}$) and the unstrained Ge channel at the Ge/$\varepsilon$-SiGe heterojunction (black points, measured at Hall density of $(4,6,8) \times 10^{10}~\mathrm{cm^{-2}}$). (h) Effective in-plane $g$-factor ($g^*_{\parallel}$) as a function of $N$ for both platforms at the same densities, extracted from the Zeeman splitting in (c), (f).}
\label{fig:four}
\end{figure*}

In both systems, the measured trends are in satisfactory agreement with our theoretical predictions based on Landau levels simulations (Supporting Information).
At a fixed density, holes confined in the Ge/$\varepsilon$-SiGe heterojunction have a larger $m^*$ and smaller $g^*_{\perp}$ compared to the $\varepsilon$-Ge quantum well, with a more pronounced sensitivity to the change in density caused by electric fields.
This behaviour arises from the reduced HH--LH energy splitting in the Ge/$\varepsilon$-SiGe heterojunction, which leads to an enhanced and density-dependent HH--LH mixing that increases at larger densities.

To extend the investigation of the electronic and spin properties of these HH--LH mixed states, we fabricated quantum point contacts (QPCs) using the same low-thermal-budget process employed for the H-FETs.
The further quantum confinement offered by these devices serves as a proxy for the future realization of quantum dots on this novel platform.
Fig.~\ref{fig:four}(a),(d) show representative atomic force microscopy (AFM) images of QPC devices realized on $\varepsilon$-Ge/SiGe quantum wells and on the same Ge/$\varepsilon$-SiGe heterojunction characterized for quantum transport.
An insulated global top accumulation gate (not shown) induces a 2DHG of density $p_{\mathrm{2D}}$, which is subsequently laterally confined into a one-dimensional channel by the two side gates.
The lithographically defined 1D channels formed by the two side gates have lateral dimensions of $\sim 300~\mathrm{nm} \times 300~\mathrm{nm}$, consistent with previous designs on Ge quantum wells \cite{hudson2025conductance}, and $\sim 200~\mathrm{nm} \times 200~\mathrm{nm}$, respectively.
The lithographically smaller channel implemented on the Ge/$\varepsilon$-SiGe heterojunction provides stronger lateral confinement, beneficial to effectively confine the expected heavier carriers.
The AFM images highlight that the vertical undulation of the cross-hatch pattern in $\varepsilon$-Ge quantum wells has a length scale comparable to the size of the QPC nanoscale gate electrodes, potentially impacting device electrostatics. Instead,this undulation is absent in the lattice-matched Ge/$\varepsilon$-SiGe platform, providing a smooth and featureless template for nanofabrication.

We observe quantized conductance plateaus as a function of side-gate voltage, indicative of ballistic transport in both material platforms, as shown in  the Supporting Information.
Source–drain bias spectroscopy of the differential transconductance $\partial G_{\mathrm{xx}}/\partial V_{\mathrm{sg}}$ as a function of the side-gate voltage $V_{\mathrm{sg}}$ and source–drain bias $V_{\mathrm{sd}}$ [Fig.~\ref{fig:four}(b–e)] reveals clear 1D subband quantization in both $\varepsilon$-Ge/SiGe and Ge/$\varepsilon$-SiGe.
The corresponding 1D subband energy spacings $E_{N,N+1}$ are extracted from these measurements by evaluating the gate
lever arm from the slopes of the transconductance diamond edges following the procedure described in \cite{hudson2025conductance} and are shown in Fig.~\ref{fig:four}(g) as a function of subband index $N$.
Also displayed are simulations of the subband energy spacings from adjusted in-plane confinement profiles and identical heterostructure parameters as in the calculation of the 2DHG confined within the heterojunction plane.
For the unstrained Ge channel at the Ge/$\varepsilon$-SiGe heterojunction, we measured three different Hall densities of $(4,6,8) \times 10^{10}~\mathrm{cm^{-2}}$ and the subband energy spacings are in good agreement with those computed from a simple parabolic model with characteristic length $\ell=28~\mathrm{nm}$ for all analysed densities.
For the $\varepsilon$-Ge quantum well, measured at a density $p_{\mathrm{2D}} = 8 \times 10^{10}~\mathrm{cm^{-2}}$, the spacings become smaller with $N$, indicating a weaker confinement strength for excited subbands.
In this case, the experimental spacings are in good agreement with a confinement profile of effective length $\ell=9.9~\mathrm{nm}$ and barrier height $V_0 = 13.5~\mathrm{meV}$ (see Supporting Information).
These values are also consistent with those reported for $\varepsilon$-Ge quantum wells on Si wafers \cite{hudson2025conductance}, which confirms the quality and reproducibility of Ge quantum point contacts.
The reduced values observed in the unstrained heterojunction reflect the expected heavier effective mass.

Zeeman spectroscopy of the QPCs [Figs.~\ref{fig:four}(c–f)], shows the corresponding evolution of spin-resolved 1D subbands with in-plane magnetic field $B$. From these measurements we evaluate the effective in-plane $g$-factor $g^*_{\parallel}$, at the same densities considered in the subband energy spectroscopy.
The in-plane $g$-factor is a key parameter for electrically driven spin-qubit operation in current hole-based quantum computing schemes, because it sets the Zeeman splitting and thus the qubit resonance condition for electric-dipole spin-resonance (EDSR) driving.
As summarized in Fig.~\ref{fig:four}(h), the unstrained Ge QPC in Ge/$\varepsilon$-SiGe exhibits higher $g^*_{\parallel}$ values compared to $\varepsilon$-Ge/SiGe, consistent with the enhanced heavy-hole--light-hole mixing discussed above.
The substantial error bars for the strained Ge quantum well $g$-factor arise because its near-zero in-plane $g$-factor produces minimal subband splitting, making the extraction uncertainty highly sensitive to intrinsic band broadening.
For this first estimate, $g^*_{\parallel}$ values are extracted assuming a zero Zeeman splitting at $B=0$ T, growing linearly as a function of in-plane magnetic field.
A quantitative agreement between experiment and theory is presented in the Supporting Information, where we account for the complex magnetic field dependence of $g^*_{\parallel}$, arising from the richer valence band structure of Ge compared to $\varepsilon$-Ge quantum wells.

\section{Conclusions}
In conclusion, we have introduced and experimentally validated a group IV semiconductor platform that hosts a high-quality buried channel in a defect-free crystalline host environment.
Being lattice-matched to the Ge substrate, our approach eliminates the need for strained relaxed buffer layers, which is promising for improving the homogeneity of future quantum dot devices built on this platform towards scalable quantum computing architectures. 
The absence of sizeable fluctuations of strain, and consequently band-offset, in the Ge/$\varepsilon$SiGe heterostructure results in a heightened susceptibility of bandstructure parameters to external electric fields, offering avenues for quantum engineering in a low-disorder, dislocation-free planar platform. 
Further tuning of the deposition parameters is expected to improve the disorder properties of the 2DHG, which already sets a benchmark for lattice-matched material stacks in group IV semiconductor, such as electrons in Si-MOS\cite{camenzind2021high,elsayed_low_2024}. 
The strong HH--LH mixing, induced by the rich valence band structure, induces in 2DHGs a tunable out-of-plane $g$-factor and in-plane effective mass, which stays light in the limit of small densities. Further confining to QPCs highlights the strong admixture of HH and LH, with smaller subband energies and larger $g^*_{\parallel}$ in Ge than in $\varepsilon$-Ge, consistent with theoretical expectations.

Unstrained Ge layers hold promise for hole spin qubits, with significantly enhanced Rabi frequencies and quality factors predicted in comparison to $\varepsilon$-Ge quantum wells \cite{bosco_squeezed_2021,secchi_hole-spin_2025,mauro_hole_2025}.
The enhanced spin-orbit coupling expected in this low-disorder system, along with the potential to host superconducting pairing correlations and the observation of fractional quantum Hall states, make this dislocation-free Ge platform promising for fast quantum hardware based on spin qubits, hybrid quantum systems based on semiconductor-superconductor quantum devices and fundamental condensed matter physics studies.

\section{Acknowledgments}

We acknowledge D.H.A.J. ten Napel, B. Morana, and the team at the Else Kooi Laboratory of TU Delft for support with the ASMI Epsilon 2000 reactor that is used for the deposition of semiconductor heterostructures. We acknowledge the research program “Materials for the Quantum Age” (QuMat) for financial support. This work was supported by the Netherlands Organisation for Scientific Research (NWO/OCW), via the Frontiers of Nanoscience program Open Competition Domain Science - M program.
We acknowledge support by the European Union through the IGNITE project with grant agreement No. 101069515 and the QLSI project with grant agreement No. 951852.
This research was sponsored in part by the Army Research Office (ARO) under Awards No. W911NF-23-1-0110. The views, conclusions, and recommendations contained in this document are those of the authors and are not necessarily endorsed nor should they be interpreted as representing the official policies, either expressed or implied, of the Army Research Office (ARO) or the U.S. Government. The U.S. Government is authorized to reproduce and distribute reprints for Government purposes notwithstanding any copyright notation herein.
This research was sponsored in part by The Netherlands Ministry of Defence under Awards No. QuBits R23/009. The views, conclusions, and recommendations contained in this document are those of the authors and are not necessarily endorsed nor should they be interpreted as representing the official policies, either expressed or implied, of The Netherlands Ministry of Defence. The Netherlands Ministry of Defence is authorized to reproduce and distribute reprints for Government purposes notwithstanding any copyright notation herein.

\section*{Data availability}
The data sets supporting the findings of this study are openly available at the Zenodo repository \cite{costa_data_2025}.

\section*{Declaration}
G.S., A.T., and L.E.A.S. are inventors on a patent application (International Application No.  PCT/NL2024/050178) submitted by Delft University of Technology related to devices in the lattice-matched Ge/SiGe heterojunction. G.S. is founding advisor of Groove Quantum BV and declares equity interests.

\end{document}


\title
  {Supporting Information for "Buried unstrained germanium channels: a lattice-matched platform for quantum technology"}
\author{Davide Costa}
\affiliation{QuTech and Kavli Institute of Nanoscience, Delft University of Technology, Lorentzweg 1, 2628 CJ Delft, Netherlands}
\author{Karina Hudson}
\affiliation{QuTech and Kavli Institute of Nanoscience, Delft University of Technology, Lorentzweg 1, 2628 CJ Delft, Netherlands}
\author{Patrick Del Vecchio}
\affiliation{QuTech and Kavli Institute of Nanoscience, Delft University of Technology, Lorentzweg 1, 2628 CJ Delft, Netherlands}
\author{Lucas E. A. Stehouwer}
\affiliation{QuTech and Kavli Institute of Nanoscience, Delft University of Technology, Lorentzweg 1, 2628 CJ Delft, Netherlands}
\author{Alberto Tosato}
\affiliation{QuTech and Kavli Institute of Nanoscience, Delft University of Technology, Lorentzweg 1, 2628 CJ Delft, Netherlands}
\author{Davide Degli Esposti}
\affiliation{QuTech and Kavli Institute of Nanoscience, Delft University of Technology, Lorentzweg 1, 2628 CJ Delft, Netherlands}
\author{Vladimir Calvi}
\affiliation{QuTech and Kavli Institute of Nanoscience, Delft University of Technology, Lorentzweg 1, 2628 CJ Delft, Netherlands}
\author{Luca Moreschini}
\affiliation{QuTech and Kavli Institute of Nanoscience, Delft University of Technology, Lorentzweg 1, 2628 CJ Delft, Netherlands}
\author{Mario Lodari}
\affiliation{QuTech and Kavli Institute of Nanoscience, Delft University of Technology, Lorentzweg 1, 2628 CJ Delft, Netherlands}
\author{Stefano Bosco}
\affiliation{QuTech and Kavli Institute of Nanoscience, Delft University of Technology, Lorentzweg 1, 2628 CJ Delft, Netherlands}
\author{Giordano Scappucci}
\email{g.scappucci@tudelft.nl}
\affiliation{QuTech and Kavli Institute of Nanoscience, Delft University of Technology, Lorentzweg 1, 2628 CJ Delft, Netherlands}

\date{\today}

\maketitle
\tableofcontents
\newpage
\section{Atomic force microscopy}

We perform a 2D atomic force microscopy (AFM) scan of the surface of the Ge/$\varepsilon$-SiGe strained-barrier heterojunction and of a reference $\varepsilon$-Ge/SiGe strained quantum well (QW) over a $16 \times 16 ~\upmu \mathrm{m}^2$ and a $20 \times 20 ~\upmu \mathrm{m}^2$ region, respectively. A 3D view of the two measurements is shown in Supplementary Fig.~\ref{fig:Sone}. The lattice-matched nature of the heterojunction is confirmed by the absence of cross-hatch pattern, which is in turn very visible on the surface of the heterostructure featuring the strained Ge QW. The extracted root mean square surface roughness values are $\sim 0.4 ~\mathrm{nm}$ and $\sim 1.8 ~\mathrm{nm}$, respectively.

\begin{figure*}[h]
  \centering
	\includegraphics[width=170mm]{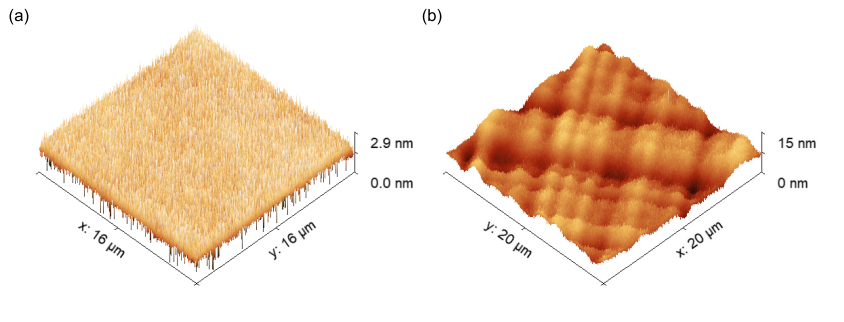}
	\caption{3D AFM of (a) the Ge/$\varepsilon$-SiGe strained-barrier heterojunction and (b) the $\varepsilon$-Ge/SiGe strained quantum well.}
\label{fig:Sone}
\end{figure*}

\newpage
\section{Raman spectroscopy}

\begin{figure*}[h]
  \centering
	\includegraphics[width=170mm]{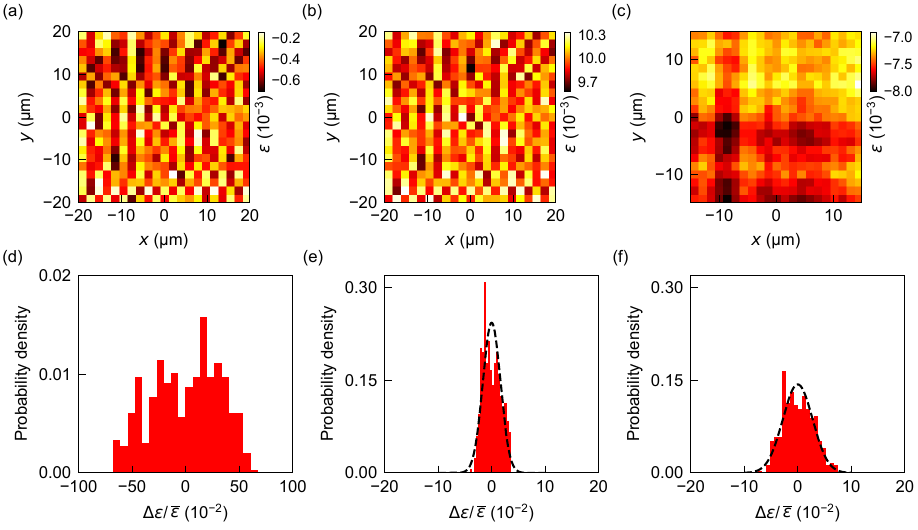}
	\caption{Raman strain maps corresponding to the $\omega_{\mathrm{Ge}}$ Raman shifts in the Ge (a) and SiGe (b) layers of the Ge/$\varepsilon$-SiGe strained-barrier heterojunction and in the Ge layer of the $\varepsilon$-Ge QW (c). (d-e-f) Strain fluctuations from the Raman maps in (a-b-c), respectively, and normal distribution fit (dashed black line). Counts are normalized such that the area under the curve integrates to one.}
\label{fig:Stwo}
\end{figure*}

We perform scanning Raman spectroscopy on both materials over a $20 \times 20 ~\upmu \mathrm{m}^2$ and a $15 \times 15 ~\upmu \mathrm{m}^2$ region, respectively. In particular, we extract the in-plane strain $\varepsilon$ from the Ge-Ge vibration $\omega_{\mathrm{Ge-Ge}}$ in the Ge (Supplementary Fig.~\ref{fig:Stwo}a) and SiGe (Supplementary Fig.~\ref{fig:Stwo}b) layers of the Ge/$\varepsilon$-SiGe strained-barrier heterojunction and in the Ge layer of the $\varepsilon$-Ge/SiGe strained QW (Supplementary Fig.~\ref{fig:Stwo}c). The mean strain values $\overline{\varepsilon}$ are $-0.42 \times 10^{-3}$ (no strain), $9.99 \times 10^{-3}$ (tensile strain) and $-7.45 \times 10^{-3}$ (compressive strain), respectively. The larger strain in the $\varepsilon$-SiGe top barrier of the heterojunction arises from the lower Ge content ($0.8$) with respect to the SiGe barrier ($0.83$), setting the lattice parameter of the $\varepsilon$-Ge quantum wells. Moreover, the strain map of the $\varepsilon$-Ge quantum well shows signatures of the cross-hatch pattern, with regions featuring higher and lower strain around $\overline{\varepsilon}$, while the strain maps of the Ge/$\varepsilon$-SiGe strained-barrier heterojunction do not. We analyse the distribution of the normalized strain fluctuations $\Delta\varepsilon/\overline{\varepsilon}$, where $\Delta\varepsilon = \varepsilon - \overline{\varepsilon}$. The strain fluctuation distributions in the $\varepsilon$-SiGe layer of the Ge/$\varepsilon$-SiGe strained-barrier heterojunction (Supplementary Fig.~\ref{fig:Stwo}e) and of the $\varepsilon$-Ge quantum well (Supplementary Fig.~\ref{fig:Stwo}f) are quite similar and can be fitted to normal distributions (dashed black lines) whereas the fluctuation of the lattice-matched Ge layer in the Ge/$\varepsilon$-SiGe strained-barrier heterojunction (Supplementary Fig.~\ref{fig:Stwo}d) have a much smaller probability density and can be therefore linked to measurement fluctuations.

\newpage
\section{Energy dispersive X-ray}

We perform an Energy Dispersive X-ray (EDX) scan of the Ge content $x_{\mathrm{Ge}}$ at the Ge/$\varepsilon$-SiGe strained-barrier heterojunction and fit it with the sigmoid function
\begin{equation}
    \frac{1}{1+e^{\frac{x-x_{0}}{\tau}}} \, ,
\end{equation}
where $x_{0}$ is the position of the interface and $\tau$ is the characteristic length quantifying the heterojunction interface. The measurement and the sigmoid fit are shown in Supplementary Fig.~\ref{fig:Sthree}, where $z = 0$ corresponds to the dielectric-semiconductor interface. We characterize the interface sharpness with the $4\tau$ parameter corresponding to the length over which $x_{\mathrm{Ge}}$ changes from $0.12$ to $0.88$ of the asymptotic value, extracting a value of $3.8(3) ~\mathrm{nm}$. Due to convolution with the EDX interaction volume, this phenomenological value represents an upper bound rather than the true physical width.
\begin{figure*}[h]
  \centering
	\includegraphics[width=85mm]{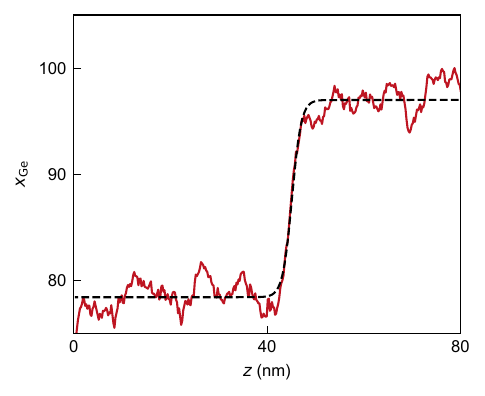}
	\caption{EDX scan of the Ge content at the Ge/$\varepsilon$-SiGe strained-barrier heterojunction with a sigmoid fit (dashed black line).}
\label{fig:Sthree}
\end{figure*}
\newpage
\section{Secondary ion mass spectroscopy}

\begin{figure*}[h]
  \centering
	\includegraphics[width=85mm]{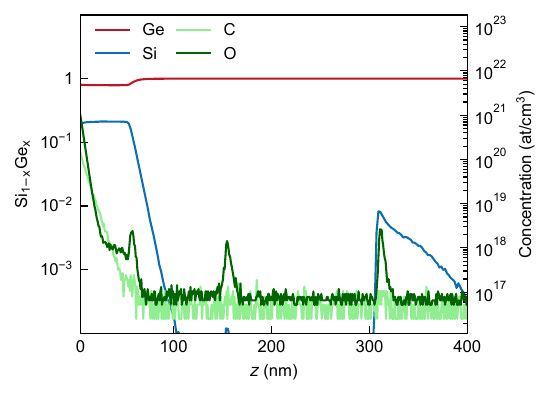}
	\caption{SIMS of the Ge/$\varepsilon$-SiGe strained-barrier heterojunction, showing Ge (red) and Si (blue) contents and O (dark green) and C (light green) concentrations.}
\label{fig:Sfour}
\end{figure*}

We analyse the chemical composition depth profile of the Ge/$\varepsilon$-SiGe strained-barrier heterojunction by secondary ion mass spectroscopy (SIMS) (Supplementary Fig.~\ref{fig:Sfour}). The measurement shows an unwanted significant oxygen accumulation at the Ge/$\varepsilon$-SiGe (2$\times$10$^{18}$ at/cm$^3$) interface about 50~nm below the surface, which may be negatively impacting the channel performance.

\newpage
\section{Experimental Methods}
\subsection{Turn-on current}

The device exhibits a clear turn-on at approximately -290 mV, corresponding to the accumulation of the 2D hole gas in the unstrained Ge channel. The vertical dashed line indicates the start of the gate voltage range analyzed to extract mobilities and densities in the main text.

\begin{figure*}[h]
  \centering
	\includegraphics[width=85mm]{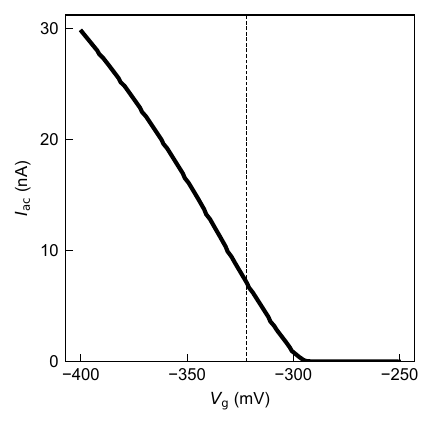}
	\caption{Source-drain AC current ($I_{\mathrm{ac}}$) measured as a function of the applied gate voltage ($V_{\mathrm{g}}$).}
\label{fig:Sfive}
\end{figure*}

\subsection{Landau fan diagram}

By continuously sweeping the applied gate voltage $V_{\mathrm{g}}$ to tune the 2D hole gas density while stepping the perpendicular magnetic field $B$ up to 6 T at a base temperature of 60 mK, we mapped the Landau fan diagram of the device. The resulting color map in Supplementary Fig.~\ref{fig:Ssix} clearly resolves the characteristic diagonal minima in $\rho_{\mathrm{xx}}$, which correspond to the formation of quantized Landau levels discussed in the main text.

\begin{figure*}[h]
  \centering
	\includegraphics[width=85mm]{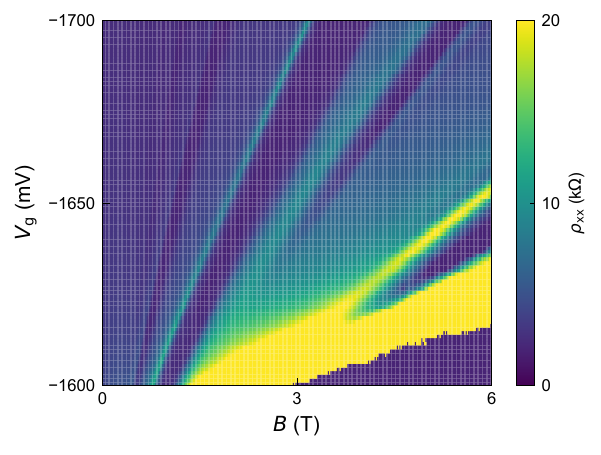}
	\caption{Color map of the longitudinal resistivity $\rho_{\mathrm{xx}}$ measured as a function of the perpendicular magnetic field $B$ and the applied gate voltage $V_{\mathrm{g}}$.}
\label{fig:Ssix}
\end{figure*}

\newpage

\subsection{2D effective mass and out-of-plane g-factor}

We extract the effective mass $m^*$ and the out-of-plane g-factor $g^*_{\perp}$ from the main text exemplary dataset in Fig.~$3$c, using magnetoresistivity $\rho_{\mathrm{xx}}(B)$ curves measured at a fixed density ($p = 4.55 \times 10^{10} ~\mathrm{cm^{-2}}$) for different $T$ (from $60$ to $850 ~\mathrm{mK}$). The activation energy gap $\Delta_{\upnu}$ of each filling factor $\nu$ can be obtained from the thermally activated decay of the Shubnikov--de Haas oscillation resistivity $\rho_{\mathrm{xx}}$ minima for a given filling factor, as reported in the Arrhenius plot of the $\ln(\rho_{\mathrm{xx}})$ against $T^{-1}$ (black circles, inset of Supplementary Fig.~\ref{fig:Sseven}). Following the Boltzmann statistics, the longitudinal magnetoresistance of a specific minima can be described via the relation $\ln(\rho_{\mathrm{xx}}) \propto -\Delta_{\upnu}/(2\mathrm{k}_{\mathrm{B}}T)$, where $\mathrm{k}_{\mathrm{B}}$ is the Boltzmann constant. Therefore, the activation energy of each filling factor can be extrapolated from the slope of a linear fit of the Arrhenius plot (dashed red line, inset of Supplementary Fig.~\ref{fig:Sseven}). Since the even and odd filling factors correspond to the cyclotron frequency and the Zeeman splitting, respectively, and a linear relation links activation energy $\Delta_{\upnu}$ and the magnetic field $B_{\upnu}$ at which each $\nu$ occurs, $m^*$ and $g^*_{\perp}$ can be extrapolated from a linear fit of the $\Delta_{\upnu}(B_{\upnu})$ dependence.
Supplementary Fig.~\ref{fig:Sseven} shows the extrapolated activation energy $\Delta_{\upnu}$ as a function of magnetic field $B$ for all the investigated even and odd filling factors (diamonds and circles, respectively). $g^*_{\perp}$ can be extrapolated from the slope of a linear fit $\Delta_{\upnu,\mathrm{odd}} = g^*_{\perp}\upmu_{\mathrm{B}}B - \Gamma$, where $\upmu_{\mathrm{B}}$ is the Bohr magneton and $\Gamma$ is the disorder-induced Landau level broadening. Once $g^*$ and $\Gamma$ are estimated, $m^*$ can be obtained from the slope of the linear fit $\Delta_{\upnu,\mathrm{even}} = \hbar e B/m^* - g^*_{\perp}\upmu_{\mathrm{B}}B - \Gamma$, where we fix $\Gamma = 134(4) ~\mathrm{\upmu eV}$ from the previous fit, due to the limited points. For the reference density $p$ of $4.55 \times 10^{10} ~\mathrm{cm^{-2}}$, we find $g^*_{\perp} = 4.0(1)$ and $m^* = 0.19(1)$.

\begin{figure*}[h]
  \centering
	\includegraphics[width=85mm]{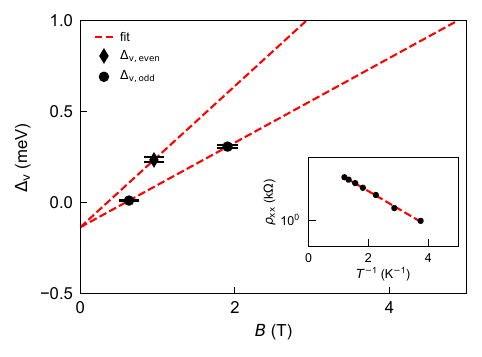}
	\caption{Activation energy gap $\Delta_{\mathrm{v}}$ as a function of magnetic field $B$ for even (diamonds) and odd (circles) filling factors $\nu$, along with linear fits (dashed red lines). We assume the two lines have the same intercept at B=0, corresponding to the disorder-induced Landau level broadening $\Gamma$. The inset shows the Arrhenius plot and fit to extract $\Delta_{\mathrm{v}}$ for $\nu = 5$.}
\label{fig:Sseven}
\end{figure*}

\newpage

\subsection{Transport metrics comparison}

\begin{figure*}[h]
  \centering
	\includegraphics[width=170mm]{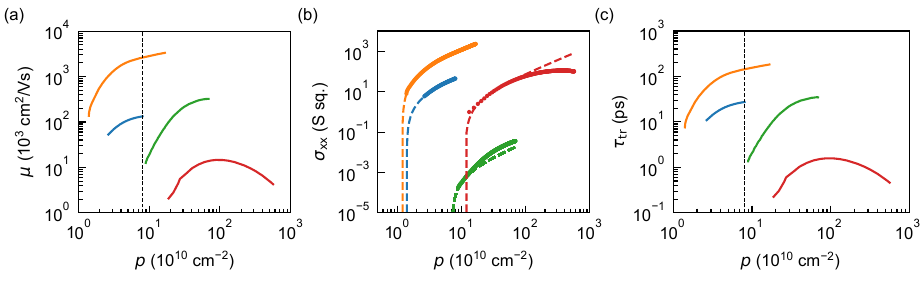}
	\caption{(a) Density-dependent mobility $\mu$, (b) longitudinal conductivity $\sigma_{\mathrm{xx}}$, and (c) transport scattering time $\tau_{\mathrm{tr}}$ measured as a function of carrier density $p$. The curves compare $\varepsilon$-Ge/SiGe quantum wells \cite{stehouwer2023germanium} (orange), the Ge/$\varepsilon$-SiGe heterojunction presented in this work (blue), $\varepsilon$-Si/SiGe quantum wells \cite{degliesposti2024low} (green), and n-type Si-MOS \cite{camenzind2021high} (red). In the middle panel, dashed lines represent fits to 2D percolation theory used to extract the critical percolation density $p_{\mathrm{p}}$ for each platform. In the left and right panels, the vertical dashed lines at $8 \times 10^{10} ~\mathrm{cm^{-2}}$ mark the carrier density at which the comparative scattering times are explicitly evaluated in the text.}
\label{fig:Seight}
\end{figure*}

Supplementary Fig.~\ref{fig:Seight} benchmarks the electrical transport properties of the 2DHG in Ge/$\varepsilon$-SiGe (this study) against 2DHGs in $\varepsilon$-Ge/SiGe \cite{stehouwer2023germanium} and 2DEGs in $\varepsilon$-Si/SiGe \cite{degliesposti2024low} and Si-MOS \cite{camenzind2021high}---state of the art material stacks that have supported functional spin-qubit devices. 

The left panel shows the density dependent mobility. We benchmark at $p=8 \times 10^{10}~\mathrm{cm^{-2}}$ (black dashed line), a convenient density at which three platforms can be directly compared and which remains relevant for quantum dot operation, corresponding to a realistic $\sim$40 nm diameter dot occupied by a single charge. At this density, holes in Ge/$\varepsilon$-SiGe achieve a mobility of $\sim$133$\times$10$^3$ cm$^2$/Vs, nearly an order of magnitude higher than electrons in $\varepsilon$-Si/SiGe ($\sim$15$\times$10$^3$ cm$^2$/Vs) \cite{degliesposti2024low}.
Relative to Si-MOS \cite{camenzind2021high}, a lattice-matched platform free of cross-hatch defects, Ge/$\varepsilon$-SiGe again shows superior performance: transport in Si-MOS is not measurable at this density, and its peak mobility ($\sim$1.5$\times$10$^4$ cm$^2$/Vs) remains an order of magnitude lower. Finally, the maximum mobility in Ge/$\varepsilon$-SiGe is much lower ($30\times$) than that of highly optimized $\varepsilon$-Ge/SiGe quantum wells ($\sim$2.5$\times$10$^6$ cm$^2$/Vs \cite{stehouwer2023germanium} at the same carrier density of 8$\times$10$^{10}$ cm$^{-2}$).

The density-dependent conductivity and percolation fits in the middle panel, however, show that both germanium platforms have similar percolation densities ($1.4(1) \times 10^{10} ~\mathrm{cm^{-2}}$ and $1.22(3) \times 10^{10} ~\mathrm{cm^{-2}}$, respectively). The percolation densities for holes in germanium-based platforms are significantly lower than the percolation densitis observed in $\varepsilon$-Si/SiGe ($6.9(1) \times 10^{10} ~\mathrm{cm^{-2}}$) and Si-MOS ($1.86 \times 10^{11} ~\mathrm{cm^{-2}}$).

In the right panel we evaluate the density-dependent transport scattering time $\tau_{\mathrm{tr}}= m_0m^*\mu/e$ to assess how differences in the effective mass $m^*$ between platforms influence the electrical performance.
For the germanium-based platforms, we used the respective density-dependent theoretical effective mass calculations $m^*(p)$ detailed in the main text, whereas for electrons in silicon we used a constant value of 0.19.
At the benchmark density of 8$\times$10$^{10}$ cm$^{-2}$ holes in Ge/$\varepsilon$-SiGe achieve a scattering time of $\sim$27 ps, which is about $5\times$ shorter than the 140 ps achieved in  $\varepsilon$-Ge/SiGe but much longer ($\sim 18\times$) than the 1.5 ps
in $\varepsilon$-Si/SiGe and Si-MOS at peak.
We speculate that the remaining factor of $\sim5\times$ difference in scattering time between Ge/$\varepsilon$-SiGe and $\varepsilon$-Ge/SiGe is due to background impurity scattering from unwanted interfacial oxygen accumulation, a well-documented effect in Ge/SiGe heterostrucutres \cite{lu2026impact}.
Notably, our measurements of interfacial oxygen concentration (2$\times$10$^{18}$ at/cm$^3$) is similar to the baseline levels reported in \cite{lu2026impact}.
Since mitigating this contamination in \cite{lu2026impact} yielded a $4\times$ increase in mobility, we envision that successfully removing this oxygen in future epitaxial iterations will enhance our platform's scattering time to make it comparable to that of state of the art  $\varepsilon$-Ge/SiGe reference.

\newpage

\subsection{1D quantization}

\begin{figure*}[h]
  \centering
	\includegraphics[width=85mm]{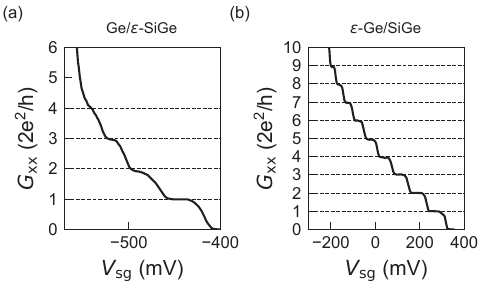}
	\caption{Longitudinal conductance $G_{\mathrm{xx}}$ measured in units of the quantum conductance $2$e$^2$/h as a function of the applied constriction gate voltage $V_{\mathrm{sg}}$. (a) Conductance trace for the Ge/$\varepsilon$-SiGe heterojunction, exhibiting distinct plateaus indicative of ballistic 1D transport. (b) Corresponding conductance trace for the $\varepsilon$-Ge/SiGe quantum well, showing well-resolved quantization steps up to higher 1D subband indices. In both panels, the horizontal dashed lines denote integer multiples of $2$e$^2$/h.}
\label{fig:Snine}
\end{figure*}

Supplementary Fig.~\ref{fig:Snine} displays the longitudinal conductance $G_{\mathrm{xx}}$ as a function of the applied split-gate voltage $V_{\mathrm{sg}}$. Panel (a) shows the transport characteristics of our unstrained Ge/$\varepsilon$-SiGe heterojunction, exhibiting quantized steps for the lowest 1D subbands.
For comparison, panel (b) displays the corresponding trace for the $\varepsilon$-Ge/SiGe quantum well, which resolves quantization steps up to higher subband indices.
As the split gates are progressively energized to deplete the underlying 2D hole gas and electrostatically narrow the conduction channel, we observe distinct quantization plateaus at integer multiples of the quantum conductance $2$e$^2$/h, as direct evidence of ballistic 1D transport.

\subsection{1D in-plane g-factor}

The effective in-plane $g$-factor $g^*_{\parallel}$ was extracted from the evolution of spin-resolved 1D subbands under an applied in-plane magnetic field $B$. 
For each subband, the Zeeman splitting $\Delta E_Z$ was obtained from the separation in side-gate voltage $V_{\mathrm{sg}}$ between the corresponding transconductance peaks as a function of magnetic field $B$, converted into energy using the subband lever arm extracted from the source-drain bias spectroscopy.
The effective $g$-factor was then obtained from the slope of the linear region of $\Delta E_Z(B)$ according to $g^*_{\parallel} = \Delta E_Z / \mu_{\mathrm{B}} B$, where $\mu_{\mathrm{B}}$ is the Bohr magneton. 
The uncertainty was evaluated from the fitting error of the transconductance peak positions.

\begin{figure*}[h]
  \centering
	\includegraphics[width=170mm]{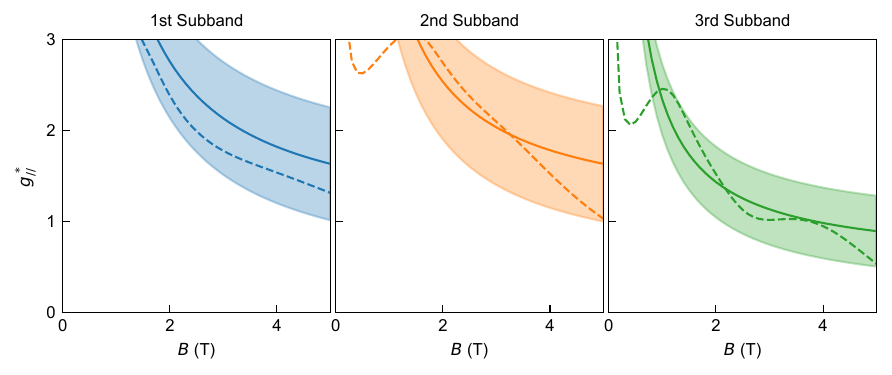}
	\caption{Experimental in-plane $g$-factor $g^*_{\parallel}$ (solid lines) and error (shaded regions) as a function of in-plane magnetic field $B$ compared with theoretical simulations (dashed lines) for the first three 1D subbands. The measured data show a non-linear dependence on $B$, consistent with the field-dependent mixing of heavy-hole and light-hole states captured by the theoretical model.}
\label{fig:Sten}
\end{figure*}

\section{Theoretical Model}

\subsection{Two-dimensional hole gas}

We consider a two-dimensional hole gas (2DHG) in a Ge/SiGe heterostructure under the influence of a magnetic $B$ and an electric field $F_z$, both applied to the $z$ direction perpendicular to the 2-dimensional plane. We compare $\varepsilon$-Ge quantum wells and Ge/$\varepsilon$-SiGe strained-barrier heterojunctions with an unstrained Ge channel.

We use $6$-band $k\cdot p$ theory to compute the energy levels and wavefunctions of the system. Because of quantum confinement in the 2DHG, we find different subbands (which we label by $j$) having a pseudospin degree of freedom (labelled by $\sigma=\pm 1$). To each of these levels is associated a 3-component spinor with envelope function components $f_j^\nu(z)\equiv\braket{z|f_j^\nu}$, where $\nu=\{\ell,s,h\}$ refer to LH, split-off hole and HH bands, respectively. For quantum wells grown on $[001]$-oriented substrates, and at $k_x=k_y=0$ and $B = 0$, the eigenstates of the system are solutions of the following Hamiltonian~\cite{DelVecchio2024}:
\begin{subequations}
  \begin{align}\label{H0}
    H_0^\mathrm{2D} &= \begin{bmatrix}
      H_{\sigma=+} & 0 \\
      0 & H_{\sigma=-}
    \end{bmatrix}, \\
    H_\sigma &= H_\sigma^k + H_\sigma^\varepsilon + V,
  \end{align}
\end{subequations}
\noindent where
\begin{subequations}
  \begin{align}
    H_\sigma^k &= \alpha_0\begin{bmatrix}
    -k_z\gamma_+k_z & 2\sqrt{2}\sigma k_z\gamma_2k_z & 0 \\
    & -k_z\gamma_1 k_z & 0 \\
    \dagger & & -k_z\gamma_-k_z
    \end{bmatrix}, \\
    H_\sigma^\varepsilon &= a_v\text{Tr}\,\varepsilon + b\cdot\delta\varepsilon\begin{bmatrix}
    -1 & \sqrt{2}\sigma & 0 \\
    & 0 & 0 \\
    \dagger & & 1
    \end{bmatrix}, \\
    V &= \mathcal{E}_{\Gamma_5^+} + \frac{\Delta_0}{3} + eF_zz - \Delta_0\begin{bmatrix}
    0 & 0 & 0 \\
    & 1 & 0 \\
    \dagger & & 0
    \end{bmatrix} .
  \end{align}
\end{subequations}
\noindent Here, $\alpha_0=\hbar^2/2m_0$, $\gamma_\pm = \gamma_1\pm 2\gamma_2$ are Luttinger parameters, $a_v$ and $b$ are strain deformation potentials, $\delta\varepsilon = \varepsilon_{xx} - \varepsilon_{zz} \approx 1.67\varepsilon_{xx}$, where $\varepsilon_{ij}$ is the strain tensor, $\Delta_0$ is the bulk split-off gap, and $\mathcal{E}_{\Gamma_5^+}$ is the valence band edge energy without spin-orbit coupling. We consider $F_z=0.1\,\text{V}/\upmu\text{m}$ ($F_z=0.05\,\text{V}/\upmu\text{m}$) and $\varepsilon_{xx}=-0.857\%$ ($\varepsilon_{xx}=-0.018\%$) for the quantum well (heterojunction). The strain in the SiGe barriers is calculated assuming pseudomorphic growth. The energy band offsets between Ge and SiGe and the deformation potential constant $b$ are obtained by linearly interpolating the values reported in reference~\cite{VandeWalle1986}. The $\gamma_i$ Luttinger parameters and $g$-factors are taken from reference~\cite{Winkler2003}. The three $\gamma_i$ and the $\kappa$ parameter are interpolated in the full composition range according to the method outlined in reference~\cite{Winkler1996}.
We stress that each material parameter ($\gamma_1$, $a_v$, $\mathcal{E}_{\Gamma_5^+}$, \dots) is a function of the position $z$ in the heterostructure. Because of this spatial dependence,  these parameters do not commute with $k_z$ and we treat them as diagonal operators in  Eq.~\eqref{H0}.

Focusing on the low-energy dynamics, the hole eigenvectors of the Hamiltonian $H_0$ in Eq.~\eqref{H0} are either of pseudospin $\sigma=\pm3/2$ ($\h{}$ levels) or pseudospin $\sigma=\pm 1/2$ ($\e{}$ levels) and they are explicitly given by
\begin{subequations}
  \begin{align}
    \ket{\h{\sigma};j} &= \Ket{\frac{3}{2},\frac{3\sigma}{2}}\Ket{f_j^h}, \\
    \ket{\e{\sigma};j} &= \Ket{\frac{3}{2},\frac{\sigma}{2}}\Ket{f_j^\ell} + \sigma\Ket{\frac{1}{2},\frac{\sigma}{2}}\Ket{f_j^s},
  \end{align}
\end{subequations}

\noindent with corresponding $\sigma$-independent energies $E_j^\h{}$ and $E_j^\e{}$, respectively. These eigenvectors comprise the bulk Bloch states $\ket{J,M_J}$ ($J=3/2, 1/2$ and $M_J= 3\sigma/2, \sigma /2$) and spatially dependent smooth envelope functions $\Ket{f_j^h} $,  $\Ket{f_j^l} $,  $\Ket{f_j^s} $.
To compute these functions, we start from the spin up ($\sigma=+1$) block $H_+$ in \eqref{H0}, which we diagonalize by finite differences methods using a $z$-mesh spacing of $0.01\,\text{nm}$ and sharp interfaces between different material systems. The eigenstates of $H_-$ are the time-reversal conjugates of the eigenstates of $H_+$.

Away from the $\Gamma$ point (with $k_x=k_y=0$), and including a finite out-of-plane magnetic field $\mathbf{B} = B\mathbf{e}_z$, the eigenstates of \eqref{H0} provide an orthonormal basis onto which the full $k\cdot p$ Hamiltonian can be projected. This results in (with bold characters indicating matrices expressed in the eigenbasis of $H_0$ and $\textbf{K}= \textbf{k}+e\textbf{A}/\hbar$ is the dynamical momentum)
\begin{equation}\label{HQW}
  \mathbf{H}^\mathrm{2D} = \mathbf{E}_0^\mathrm{2D} + \alpha_0\mathbf{M}_\gamma K_\parallel^2 + \frac{\alpha_0}{2l_B^2}\mathbf{M}_g + \alpha_0\left(i\mathbf{M}_1 K_- + \mathbf{M}_2 K_-^2 + \text{h.c.}\right),
\end{equation}
\noindent where $K_\pm = K_x \pm iK_y$, $K_\parallel^2 = K_x^2 + K_y^2 = \{K_-,K_+\}/2$ , and
\begin{equation}
  \mathbf{E}_0^\mathrm{2D} = \begin{bmatrix}
    \mathbf{E}^\h{} & 0 & 0 & 0 \\
    & \mathbf{E}^\e{} & 0 & 0 \\
    & & \mathbf{E}^\e{} & 0 \\
    & & & \mathbf{E}^\h{}
  \end{bmatrix},
\end{equation}

\noindent with $\mathbf{E}^\tau = \text{diag}\{E_1^\tau,E_2^\tau,\dots\}$ ($\tau=\{\e{},\h{}\}$) are the energies from \eqref{H0}. The $\mathbf{M}$-matrices are
\begin{subequations}\label{Mmatrices}
  \begin{align}
    \mathbf{M}_\gamma &= \begin{bmatrix}
      \boldsymbol{\Gamma}_\parallel^\h{} & 0 & 0 & 0 \\
      & \boldsymbol{\Gamma}_\parallel^\e{} & 0 & 0 \\
      & & \boldsymbol{\Gamma}_\parallel^\e{} & 0 \\
      & & & \boldsymbol{\Gamma}_\parallel^\h{}
    \end{bmatrix},
    & \mathbf{M}_g &= \begin{bmatrix}
      \mathbf{G}_\perp^\h{} & 0 & 0 & 0 \\
      & \mathbf{G}_\perp^\e{} & 0 & 0 \\
      & & -\mathbf{G}_\perp^\e{} & 0 \\
      & & & -\mathbf{G}_\perp^\h{}
    \end{bmatrix}, \\
    \mathbf{M}_1 &= \begin{bmatrix}
      0 & \mathbf{T}^\text{x} & 0 & 0 \\
      0 & 0 & \mathbf{T}^\e{} & 0 \\
      0 & 0 & 0 & \mathbf{T}^{\text{x}\dagger} \\
      \mathbf{T}^\h{} & 0 & 0 & 0
    \end{bmatrix},
    & \mathbf{M}_2 &= \begin{bmatrix}
      0 & 0 & \boldsymbol{\upmu} & 0 \\
      0 & 0 & 0 & \boldsymbol{\upmu}^\dagger \\
      \boldsymbol{\updelta}^\dagger & 0 & 0 & 0 \\
      0 & \boldsymbol{\updelta} & 0 & 0
    \end{bmatrix}.
  \end{align}
\end{subequations}
These matrix elements are explicitly expanded in terms of the eigenstates of \eqref{H0} as:
\begin{subequations}
  \begin{align}
\boldsymbol{\Gamma}_\parallel^\h{} &= -\braket{f^h|\gamma_1 + \gamma_2|f^h}, \\
    \mathbf{G}_\perp^\h{} &= -\braket{f^h|6\kappa + \frac{27q}{2}|f^h}, \\
    \boldsymbol{\Gamma}_\parallel^\e{} &= -\braket{f^z|\gamma_-|f^z} - \braket{f^\circ|\gamma_1 + \gamma_2|f^\circ}, \\
    \mathbf{G}_\perp^\e{} &= -6\braket{f^\circ|\kappa|f^\circ} - \frac{1}{2}\braket{f^\ell|q|f^\ell} + 2\left(\braket{f^z|f^z} - 2\braket{f^\circ|f^\circ}\right), \\
    \mathbf{T}^\text{x} &= -\frac{3i}{\sqrt{2}}\bra{f^h}\left(u_+\ket{f^z} + \frac{7\sqrt{6}}{12}[q,k_z]\ket{f^\ell}\right), \\
    \mathbf{T}^\h{} &= -\frac{3i}{2}\braket{f^h|[q,k_z]|f^h}, \\
    \mathbf{T}^\e{} &= -\frac{3i}{\sqrt{2}}\left(\braket{f^\circ|u_+|f^z} - \braket{f^z|u_-|f^\circ}\right) - 5i\braket{f^\ell|[q,k_z]|f^\ell}, \\
    \boldsymbol{\upmu} &= \frac{3}{2}\braket{f^h|\gamma_2 + \gamma_3|f^\circ}, \\
    \boldsymbol{\updelta} &= \frac{3}{2}\braket{f^h|\gamma_2 - \gamma_3|f^\circ},
  \end{align}
\end{subequations}
\noindent where $u_\pm = \{\gamma_3,k_z\} \pm [\kappa,k_z]$, $\{A,B\} \equiv AB + BA$ is the anti-commutator and 
\begin{subequations}
  \begin{align}
    \ket{f_j^z} &\equiv \frac{1}{\sqrt{3}}\left(\sqrt{2}\ket{f_j^\ell} - \ket{f_j^s}\right), \\
    \ket{f_j^\circ} &\equiv \frac{1}{\sqrt{3}}\left(\ket{f_j^\ell} + \sqrt{2}\ket{f_j^s}\right).
  \end{align}
\end{subequations}
\noindent We omit for simplicity the explicit subband indices.

\subsection{Landau levels}
At large magnetic field values, we simulate the Landau level eigenspectrum as a function of the magnetic field applied out-of-plane.

 In this case,   the momenta in $x$ and $y$ direction do not commute and $[K_x,K_y]=-i\text{sign}(B)/l_B^2$, with $l_B=\sqrt{\hbar/e|B|}\sim 26$~nm at $B=1$~T. 
Here,  we consider positive values of $B$, and we define the dimensionless Landau level operators $a=iK_- l_B/\sqrt{2}$ and $a^\dagger=-iK_+ l_B/\sqrt{2}$, satisfying $[a,a^\dagger]=1$. Note that the system has an additional degrees of freedom, the relative center of mass coordinate, which  does not enter the Hamiltonian and thus produces the well-known degeneracy $\mathcal{N}=L_x L_y/2\pi l_B^2=\phi/\phi_0$, coming by assuming the periodic boundary conditions in $x$ and $y$ directions. Here, $\phi$ is the magnetic flux in the sample of area $L_x L_y$ and $\phi_0$ is the flux quantum.

We note that $K_\parallel^2= (2a^\dagger a+1)/l_B^2$, and that $\hbar^2/m_0 l_B^2= 2\mu_B B$, such that the Hamiltonian in Eq.~\eqref{HQW} reduces to 
\begin{equation}\label{Hfan}
  \begin{split}
    \tilde{\mathbf{H}} &= \begin{bmatrix}
      \mathbf{E}_+^\h{}(B) & & & \\
      & \mathbf{E}_+^\e{}(B) & & \\
      & & \mathbf{E}_-^\e{}(B) & \\
      & & & \mathbf{E}_-^\h{}(B)
    \end{bmatrix} \\
    &+ \frac{\alpha_0}{l_B^2}\begin{bmatrix}
      2\boldsymbol{\Gamma}_\parallel^\h{}a^\dag a & \sqrt{2}l_B\mathbf{T}^\text{x}a & -2\boldsymbol{\upmu}a^2 & \mathbf{0} \\
      & 2\boldsymbol{\Gamma}_\parallel^\e{}a^\dag a & \sqrt{2}l_B\mathbf{T}^\e{}a & -2\boldsymbol{\upmu}^\dag a^2 \\
      \dag & & 2\boldsymbol{\Gamma}_\parallel^\e{}a^\dag a & \sqrt{2}l_B\mathbf{T}^{\text{x}\dag}a \\
      & & & 2\boldsymbol{\Gamma}_\parallel^\h{}a^\dag a
    \end{bmatrix},
  \end{split}
\end{equation}
\noindent where $\mathbf{E}_\pm^\tau(B) = \mathbf{E}^\tau + \frac{\alpha_0}{l_B^2}\left(\boldsymbol{\Gamma}_\parallel^\tau \pm \mathbf{G}_\perp^\tau/2\right)$ and we use the axial approximation $\gamma_2=\gamma_3$. The distribution of the various powers of $a$ and $a^\dagger$ in the Hamiltonian \eqref{Hfan} suggests that all its eigenvectors $\ket{\varphi}$ can be written as

\begin{equation}\label{Hfanphi}
  \ket{\varphi} = \begin{bmatrix}
    \mathbf{c}_{\h{}+}\ket{m-1} \\
    \mathbf{c}_{\e{}+}\ket{m} \\
    \mathbf{c}_{\e{}-}\ket{m+1} \\
    \mathbf{c}_{\h{}-}\ket{m+2}
  \end{bmatrix},
\end{equation}
\noindent where the $\mathbf{c}_{\tau\pm}$ are scalars (one scalar for each $\h{+}$ subband, and so on) and the $\ket{m}$ are the Fock states of the Landau levels operators, i.e. $a^\dagger a\ket{m} = \ket{m}m$. By projecting  the Hamiltonian \eqref{Hfan} onto the eigenvectors \eqref{Hfanphi}, we find a block diagonal Hamiltonian that involves only the integer $m$: 
\begin{equation}\label{HLL}
  \begin{split}
    \tilde{\mathbf{H}}^{m\geq 1} &= \begin{bmatrix}
      \mathbf{E}_+^\h{}(B) & & & \\
      & \mathbf{E}_+^\e{}(B) & & \\
      & & \mathbf{E}_-^\e{}(B) & \\
      & & & \mathbf{E}_-^\h{}(B)
    \end{bmatrix} \\
    &+ \frac{\alpha_0}{l_B^2}\begin{bmatrix}
      2\boldsymbol{\Gamma}_\parallel^\h{}(m-1) & \sqrt{2}l_B\mathbf{T}^\text{x}\sqrt{m} & -2\boldsymbol{\upmu}\sqrt{m(m+1)} & \mathbf{0} \\
      & 2\boldsymbol{\Gamma}_\parallel^\e{}m & \sqrt{2}l_B\mathbf{T}^\e{}\sqrt{m+1} & -2\boldsymbol{\upmu}^\dagger\sqrt{(m+1)(m+2)} \\
      & & 2\boldsymbol{\Gamma}_\parallel^\e{}(m+1) & \sqrt{2}l_B\mathbf{T}^{\text{x}\dagger}\sqrt{m+2} \\
      & & & 2\boldsymbol{\Gamma}_\parallel^\h{}(m+2)
    \end{bmatrix}. 
  \end{split}
\end{equation}
\noindent This Hamiltonian can be diagonalized to extract the set of scalars $\mathbf{c}_{\tau\sigma}$ for any $m\geq -2$. For $m=-2$, only the $\h{-}$ part of $\ket{\varphi}$ is well defined (since $\ket{m+2}=\ket{0}$), and implies that only the scalars $\mathbf{c}_{\h{-}}$ are non-zero. For $m=-1$, the $\h{-}$ as well as the $\e{-}$ subspaces are well-defined, and only $\mathbf{c}_{\h{-}}$ and $\mathbf{c}_{\e{-}}$ are non-zero. Similarly, the three subspaces $\h{-}$, $\e{-}$ and $\e{+}$ are well defined for $m=0$, and therefore $\mathbf{c}_{\h{+}}$ must be zero. For any $m\geq 1$, the Hamiltonian \eqref{HLL} is diagonalized directly without any conditions on $\mathbf{c}_{\tau\sigma}$. We remark that this procedure is independent of the gauge.

We truncate the system to include only the levels from $m=-2$ to $m=11$ and vary $B$. This gives a total of $2\times 11 + 3=25$ magnetic field-dependent spin-polarized Landau levels for each HH subband in the $z$-direction, which we now label by $n\geq 1$, each one degenerate by $\mathcal{N}$ because of the residual center of mass degree of freedom.

The dependence on $B$ of the numerically computed Landau level energies $\epsilon_{n}$ (labelled in increasing order of energy) enables us to extract the $B$-field-dependent effective spin-resolved mass and g factor of the hole gas as 
\begin{equation}
\label{eq:mandg}
\frac{1}{m^*_{\downarrow}} = \frac{\epsilon_{n=3}-\epsilon_{n=1}}{2\mu_B B} \ , \ \ \ \
\frac{1}{m^*_{\uparrow}} = \frac{\epsilon_{n=4}-\epsilon_{n=2}}{2\mu_B B} \ , \  \text{and}\  \
g_\perp^* = \frac{\epsilon_{n=2}-\epsilon_{n=1}}{\mu_B B}\ . 
\end{equation}
The density $p$ is related to the magnetic field via the filling factor $\nu=2\pi l_B^2 p$, which then results in $B=b_0 p/\nu$, with $b_0=h/e=4.13 \ \text{T}/(10^{11} \ \text{cm}^{-2})$. This definition of $g$ factor and effective mass is consistent with thermal activation energy measurements, probing the activation energy of the $\nu=1$ ($\nu=2$) filling factor for $m^*_{\downarrow}$  and $g_\perp^*$ ($m^*_{\uparrow}$). The plots in the main text display the extracted spin down mass $m^*_{\downarrow}$ and $g$-factor obtained from \eqref{eq:mandg}. We exclude from the plots the additional levels with $n>4$.

\subsection{1D channels}

We model strained and unstrained Ge QPCs by introducing an in-plane confinement profile $V_\parallel(x)$ to the $k\cdot p$ Hamiltonian, where $x$ is the direction of quantization and $y$ is parallel to the direction of motion. We describe magnetic fields directed towards the $y$-axis with the Landau gauge
\begin{equation}
    \mathbf{A}(\mathbf{r}) = -Bx\mathbf{e}_z,
\end{equation}

\noindent such that $\nabla\times\mathbf{A} = B\mathbf{e}_y$. In the unstrained Ge QPC the measured subband energy splitting is relatively constant, indicating that an harmonic confinement suffices to describe the data and avoids over-fitting. Therefore, in this case, we use a simple parabolic confinement $V^\mathrm{u}_\parallel(x) = -\alpha_0x^2/\ell^4$ with $\ell = 28\,\mathrm{nm}$ and where the minus sign comes from our chosen hole-energy convention. In the strained Ge QPC, where the measured subband gap is largely anharmonic, we model the confinement with
\begin{equation}
    V_\parallel^\mathrm{s}(x) = -V_0\left[1-\frac{1}{\sqrt{1+2\alpha_0x^2/(V_0\ell^4)}}\right],
\end{equation}

\noindent where $\ell = 9.9\,\mathrm{nm}$ and $V_0 = 13.5\,\mathrm{meV}$. This confinement potential nicely reproduces the $\sim 1/n$ trend of the higher energy gaps. Note that $V_\parallel^\mathrm{s}(|x|\to\infty) = -V_0$ and $V_\parallel^\mathrm{s}(x) \simeq -\alpha_0x^2/\ell^4$ when $x\ll \ell^2\sqrt{V_0/(2\alpha_0)}$. We remark that $\alpha_0 = \hbar^2/(2m_0)$, i.e. the effective parabolic length $\ell$ is related to that of a free electron of mass $m_0$. The strained Ge QPC Hamiltonian is solved in position space by means of finite differences with $k_x = -i\partial_x$, a mesh step $\delta x = 1\,\mathrm{nm}$ and in the domain $x\in [-750, +750]\,\mathrm{nm}$. The unstrained Ge QPC Hamiltonian is projected onto the first $30$ eigenstates of the harmonic oscillator, obtained by writing $x = \frac{\ell}{\sqrt{2}}\left(a + a^\dagger\right)$ and $k_x = \frac{1}{\sqrt{2}i\ell}\left(a - a^\dagger\right)$, where $a$ is the usual ladder operator.

We first diagonalize the 1D channel Hamiltonian $\mathbf{H}_0^\mathrm{1D}$ at $B=0$ and $k_y = 0$. This later provides an orthonormal basis on which the full Hamiltonian at finite $B$ and $k_y$ is projected. This Hamiltonian is 
\begin{equation}\label{H1D_0}
    \mathbf{H}_0^\mathrm{1D} = \mathbf{E}_0^\mathrm{2D} + \alpha_0\left[\left(\mathbf{M}_\gamma + \mathbf{M}_2 + \mathbf{M}_2^\dagger\right) k_x^2 + i\left(\mathbf{M}_1 - \mathbf{M}_1^\dagger\right)k_x\right] + V_\parallel(x),
\end{equation}

\begin{figure}
\centering
\includegraphics[scale=1]{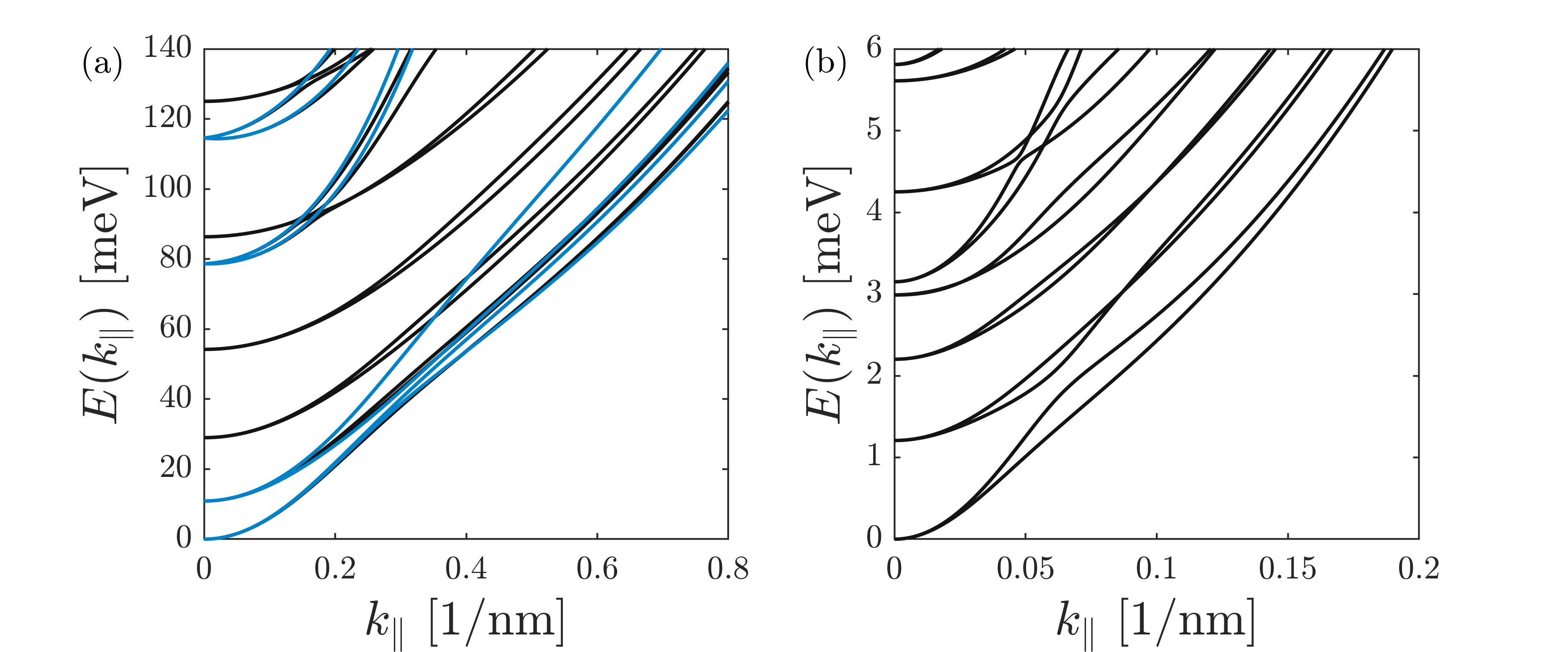}
\caption{Energy dispersion $E(k_\parallel)$ of the (a) strained Ge QW computed from the full 200-subband matrix (black) and from the effective 4-band matrix (blue) and (b) of the unstrained Ge system  computed from the 200-subband matrix.}\label{fig:bs}
\end{figure}

\noindent where $\mathbf{E}_0^\mathrm{2D}$ are the energies of the planar system without in-plane confinement. For unstrained Ge, the $\mathbf{M}$-matrices correspond to those defined in Eq.~\eqref{Mmatrices}, truncated to include the first $200$ 2D subbands. This results in a Hamiltonian of dimension $2\times 30\times 200 = 12000$. In strained Ge, given the large basis along the $x$-dimension, we collapse the 2D basis down onto the first $2$ HH levels $+$ the first $2$ LH levels, with the remaining $200-4 = 196$ subbands are included by a $2\textsuperscript{nd}$ order perturbation. The ground state dispersion from the 4-subband Hamiltonian is in good agreement with that from the complete 200-subband Hamiltonian over a range of wavenumber that is well beyond the requirements of the following calculations ($\sim 1/(10\,\mathrm{nm}) = 0.1\,\mathrm{nm}^{-1}$, see Supplementary Fig. \ref{fig:bs}). The resulting 1D Hamiltonian is of dimension $2\times 1501\times 4 = 12008$ and has the same structure as \eqref{H1D_0}, but with renormalized $\mathbf{M}$-matrices:

\begin{subequations}
    \begin{align}
    \tilde{\mathbf{M}}_1 &= \mathbf{M}_1, \\
    \tilde{M}^i_{\gamma,j} &= M^i_{\gamma,j} + \alpha_0\sum_{k\in\mathcal{B}}\left(\frac{M^i_{1,k}M^{\dagger k}_{1,j}}{E_i - E_k} + \frac{M^i_{1,k}M^{\dagger k}_{1,j}}{E_j - E_k} + \frac{M^{\dagger i}_{1,k}M^k_{1,j}}{E_i - E_k} + \frac{M^{\dagger i}_{1,k}M^k_{1,j}}{E_j - E_k}\right),  \\
    \tilde{M}^i_{2,j} &= M^i_{2,j} - \alpha_0\sum_{k\in\mathcal{B}}\left(\frac{M^i_{1,k}M^k_{1,j}}{E_i - E_k} + \frac{M^i_{1,k}M^k_{1,j}}{E_j - E_k}\right),
\end{align}
\end{subequations}

\noindent where the $\mathcal{B}$-set refers to the remote $196$ subbands. The Hamiltonian $\mathbf{H}_0^\mathrm{1D}$ is diagonalized for the first $501$ channel subbands for unstrained Ge and for the first $540$ subbands for strained Ge (not counting spin in both cases). Then the full 1D channel Hamiltonian is projected onto the 1D channel basis states, and diagonalized at finite $B$ and $k_y$. The result is

\begin{equation}
    \mathbf{H}^\mathrm{1D} = \mathbf{E}_0^\mathrm{1D} + \alpha_0\left(\mathbf{L}_1k_y + \mathbf{L}_\gamma k_y^2 + \frac{1}{2l_B^2}\mathbf{L}_{g\parallel} + \frac{1}{2l_B^2}\mathbf{L}_{3\parallel}k_y + \frac{1}{4l_B^4}\mathbf{L}_{4\parallel}\right),
\end{equation}

\noindent where the $\mathbf{L}$-matrices are

\begin{subequations}
    \begin{align}
    L^i_{1,j} &= v^{\dagger i}{}_k\left[m^k_{1,l} - 2(\mathbf{m}_2k_x)^k{}_l\right]v^l{}_j, \\
    L^i_{\gamma,j} &= v^{\dagger i}{}_k\left[m^k_{0,l}\right]v^l{}_j, \\
    L^i_{g\parallel,j} &= v^{\dagger i}{}_k\left[n^k_{g,l} - (\mathbf{N}^\prime_\gamma x)^k{}_l + (\mathbf{n}^\prime_1\{x,k_x\})^k{}_l\right]v^l{}_j, \\
    L^i_{3\parallel,j} &= v^{\dagger i}{}_k\left[2(\mathbf{n}_1x)^k{}_l\right]v^l{}_j, \\
    L^i_{4\parallel,j} &= v^{\dagger i}{}_k\left[(\mathbf{N}_\gamma x^2)^k{}_l\right]v^l{}_j, \\
\end{align}
\end{subequations}

\noindent with $v^i{}_j$ being the $i$-th component of the $j$-th eigenvector of $\mathbf{H}^\mathrm{1D}_0$ (i.e. $v^{\dagger i}{}_kv^k{}_j = \delta^i_j$) and where

\begin{subequations}
    \begin{align}
    \mathbf{m}_0 &= \mathbf{M}_\gamma - \mathbf{M}_2 - \mathbf{M}_2^\dagger, \\
    \mathbf{m}_1 &= \mathbf{M}_1 + \mathbf{M}_1^\dagger, \\
    \mathbf{m}_2 &= i\left(\mathbf{M}_2 - \mathbf{M}_2^\dagger\right), \\
    \mathbf{n}_g &= -i\left(\mathbf{N}_g - \mathbf{N}_g^\dagger\right), \\
    \mathbf{n}_1 &= \mathbf{N}_1 + \mathbf{N}_1^\dagger, \\
    \mathbf{n}^\prime_1 &= i\left(\mathbf{N}_1 - \mathbf{N}_1^\dagger\right).
\end{align}
\end{subequations}

The $\mathbf{N}$-matrices describe the 2D system with in-plane magnetic fields:

\begin{align}
    \mathbf{N}_g &= \begin{bmatrix}
        0 & \mathbf{G}_\parallel^\times & 0 & 0 \\ 0 & 0 & \mathbf{G}_\parallel^\e{} & 0 \\ 0 & 0 & 0 & \mathbf{G}_\parallel^{\times\dagger} \\ 0 & 0 & 0 & 0
    \end{bmatrix}, &
    \mathbf{N}_1 &= \begin{bmatrix}
        0 & \mathbf{R}^\times & 0 & 0 \\ 0 & 0 & \mathbf{R}^\e{} & 0 \\ 0 & 0 & 0 & \mathbf{R}^{\times\dagger} \\ 0 & 0 & 0 & 0
    \end{bmatrix}, \\
    \mathbf{N}_\gamma &= \begin{bmatrix}
        \boldsymbol{\Gamma}_\perp^{\h{}} & 0 & 0 & 0 \\ 0 & \boldsymbol{\Gamma}_\perp^{\e{}} & 0 & 0 \\ 0 & 0 & \boldsymbol{\Gamma}_\perp^{\e{}} & 0 \\ 0 & 0 & 0 & \boldsymbol{\Gamma}_\perp^{\h{}}
    \end{bmatrix}, &
    \mathbf{N}^\prime_\gamma &= \begin{bmatrix}
    \boldsymbol{\Gamma}_\perp^{\prime\h{}} & 0 & 0 & 0 \\ 0 & \boldsymbol{\Gamma}_\perp^{\prime\e{}} & 0 & 0 \\ 0 & 0 & \boldsymbol{\Gamma}_\perp^{\prime\e{}} & 0 \\ 0 & 0 & 0 & \boldsymbol{\Gamma}_\perp^{\prime\h{}}
    \end{bmatrix},
\end{align}

\noindent where

\begin{subequations}
    \begin{align}
    \mathbf{G}_\parallel^\e{} &= -\sqrt{2}\left(\bra{f^\circ}3\kappa+1\ket{f^z} + \bra{f^z}3\kappa+1\ket{f^\circ}\right) + 2\braket{f^z|f^z}, \\
    \mathbf{G}_\parallel^\times &= -\sqrt{2}\bra{f^h}\left(3\kappa\ket{f^z} - \sqrt{3}\ket{f^s}\right), \\
    \mathbf{R}^\e{} &= 3\sqrt{2}i\left(\bra{f^\circ}\gamma_3\ket{f^z} - \bra{f^z}\gamma_3\ket{f^\circ}\right), \\
    \mathbf{R}^\times &= 3\sqrt{2}i\bra{f^h}\gamma_3\ket{f^z}, \\
    \boldsymbol{\Gamma}_\perp^\h{} &= -\bra{f^h}\gamma_1-2\gamma_2\ket{f^h}, \\
    \boldsymbol{\Gamma}_\perp^{\prime\h{}} &= -\frac{1}{2}\bra{f^h}\{\gamma_1-2\gamma_2,k_z\}\ket{f^h}, \\
    \boldsymbol{\Gamma}_\perp^\e{} &= -\bra{f^\circ}\gamma_1-2\gamma_2\ket{f^\circ}-\bra{f^z}\gamma_1+4\gamma_2\ket{f^z}, \\
    \boldsymbol{\Gamma}_\perp^{\prime\e{}} &= -\frac{1}{2}\bra{f^\circ}\{\gamma_1-2\gamma_2,k_z\}\ket{f^\circ}-\frac{1}{2}\bra{f^z}\{\gamma_1+4\gamma_2,k_z\}\ket{f^z}.
\end{align}
\end{subequations}

The Zeeman splitting as a function of $B$ is then computed for fixed $k_y = \sqrt{2\pi p}$, determined by the reported hole density for each experiment: $p=5\times 10^{10}\,\mathrm{cm}^{-2}$ in the unstrained Ge channel and $p=6\times 10^{10}\,\mathrm{cm}^{-2}$ in the strained Ge channel. The computed $g$-factor $g_\parallel = \Delta E/\mu_\mathrm{B}B$ for the unstrained Ge QPC is plotted against the $g$-factor extracted from experiment in Supplementary Fig. \ref{fig:Ssix} (dashed lines). The computed $g$-factor as a function of $B$ for the strained Ge QPC is plotted in Supplementary Fig. \ref{fig:gi-QW}.

\begin{figure}
    \centering
    \includegraphics[width=0.5\linewidth]{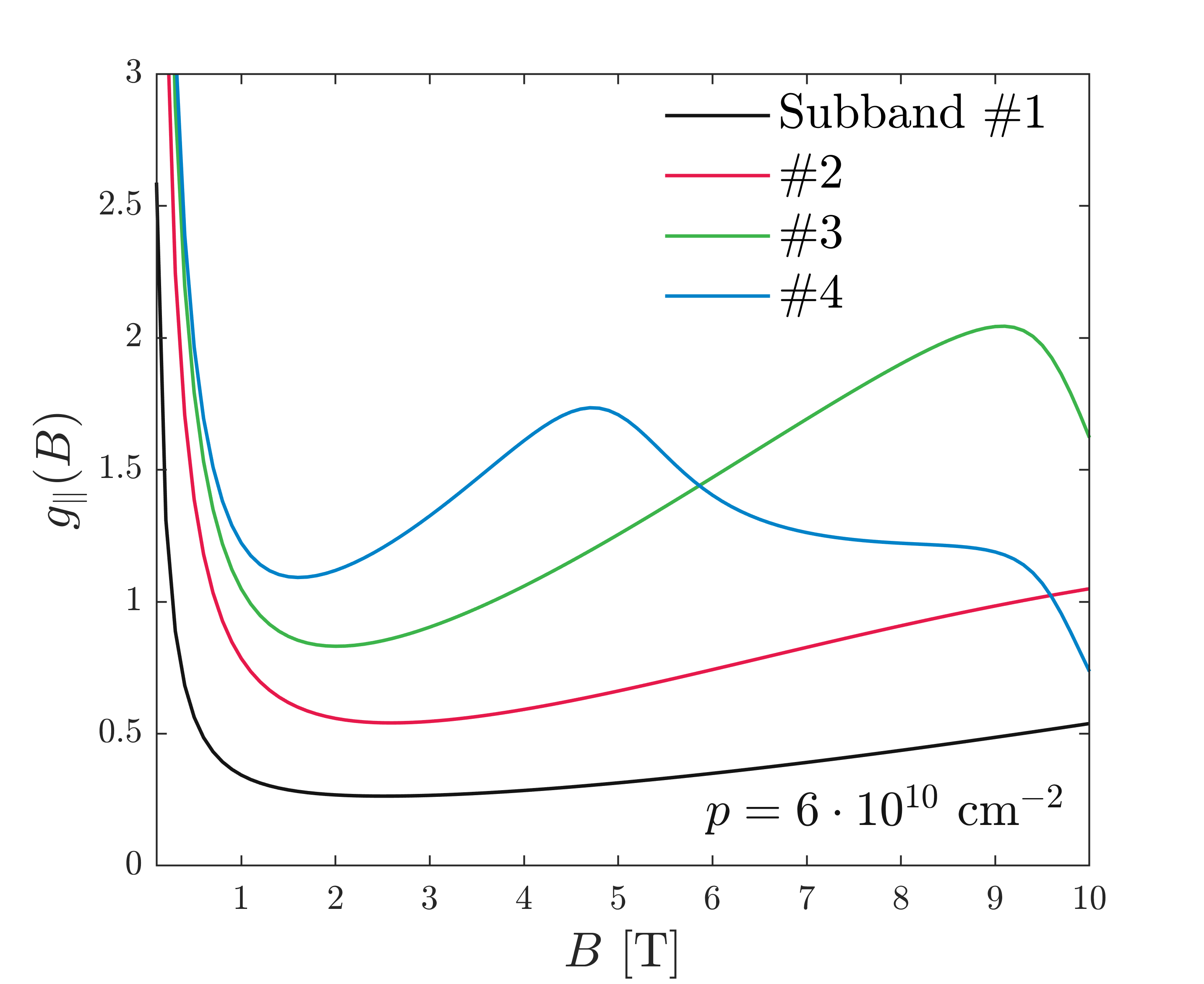}
    \caption{Calculated in-plane $g$-factor of the first $4$ 1D subbands as a function of $B$ for the strained Ge QPC.}
    \label{fig:gi-QW}
\end{figure}